%% file: main.tex
\documentclass[%
aip,
superscriptaddress,
 amsmath,amssymb,
preprint,%
]{revtex4-2}

\usepackage{graphicx}
\usepackage{dcolumn}
\usepackage{bm}
\usepackage[english]{babel}
\usepackage{blindtext}
\usepackage[utf8]{inputenc}
\usepackage[T1]{fontenc}
\usepackage{mathptmx}
\usepackage{float}
\usepackage{xcolor}
\usepackage{listings}
\usepackage{subfig}
\usepackage{booktabs}

\newcommand{\pdiff}[2]%
{
  \frac{\partial#1}{\partial#2}
}
\def\beq{\begin{equation}}
\def\eeq{\end{equation}}
\def\beqar{\begin{eqnarray}}
\def\eeqar{\end{eqnarray}}
\def\grad{\nabla}

\begin{document}



\title{INGRID: an interactive grid generator for 2D edge plasma modeling\\}

\author{B. M. Garcia}
\email{bgarci26@ucsc.edu}
\affiliation{ 
University of California - Santa Cruz, Santa Cruz, California, USA 
}

\author{M. V. Umansky}
\email{umansky1@llnl.gov}
\affiliation{
Lawrence Livermore National Laboratory, Livermore, California, USA
}

\author{J. Watkins}
\email{joeymwatkins@gmail.com}
\affiliation{ 
Brigham Young University, Idaho, USA
}

\author{J. Guterl}
\email{guterlj@fusion.gat.com}
\affiliation{ 
General Atomics, San Diego, California, USA
}

\author{O. Izacard}
\email{izacard1@llnl.gov}
\affiliation{
Lawrence Livermore National Laboratory, Livermore, California, USA
}

\date{\today}
\include{abstract}

\maketitle

\section{Introduction}
\input{introduction/tokamak_edge_plasma_modeling}

\input{introduction/grid_generation_for_tokamak}
\input{introduction/outline}

\section{Classification of equilibria}
\input{classification/all_topology}

\section{Computational algorithm methodology}
\input{computation/divide_conquer}
\input{computation/bicubic_interpolation}
\input{computation/refining_reference_points}
\input{computation/topology_analysis}
\input{computation/patch_map_construction}
\input{computation/grid_construction}
\input{computation/grid_optimization}

\section{Performance}
\input{performance/scaling_time}
\input{performance/benchmarks}

\section{Software design and user interaction}

\input{software_methodology/development_in_python}
\input{software_methodology/software_organization}
\input{software_methodology/geometry_hierarchy}
\input{software_methodology/ingrid_parameter_file}
\input{software_methodology/ingrid_target_plate_file}
\input{software_methodology/ingrid_workflow}

\section{Summary}
\input{closing/conclusion}

\section{Appendix: Parameter file}
\input{appendix}

\bibliography{ingrid_bib}
\input{figures}

\end{document}

%% file: abstract.tex
\begin{abstract}

A fusion boundary-plasma domain is defined by axisymmetric magnetic
surfaces where the geometry is often complicated by the presence of
one or more X-points; and modeling boundary plasmas usually relies on
computational grids that account for the magnetic field geometry. The
new grid generator INGRID (Interactive Grid Generator) presented here
is a Python-based code for calculating grids for fusion boundary
plasma modeling, for a variety of configurations with one or two
X-points in the domain. Based on a given geometry of the magnetic
field, INGRID first calculates a skeleton grid which consists of a
small number of quadrilateral patches; then it puts a subgrid on each
of the patches, and joins them in a global grid. This domain partitioning 
strategy makes possible a uniform treatment of various
configurations with one or two X-points in the domain.  This includes
single-null, double-null, and other configurations with two X-points
in the domain. The INGRID design allows generating grids either
interactively, via a parameter-file driven GUI, or using a
non-interactive script-controlled workflow. Results of testing
demonstrate that INGRID is a flexible, robust, and user-friendly
grid-generation tool for fusion boundary-plasma modeling.

\end{abstract}

%% file: introduction/tokamak_edge_plasma_modeling.tex
\subsection{Tokamak edge plasma modeling}
Research in tokamak edge plasma physics is critical for realizing
practical fusion energy and designing future fusion reactors. One of
the greatest challenges that tokamak edge plasma researchers face
today is determining effective methods for controlling particle and
heat fluxes on tokamak plasma-facing components (PFC) while maintaining good core-plasma performance. A possible
solution is the use of advanced divertor configurations
\cite{Kotschenreuther2007}.

The traditional X-point divertor configuration uses a first-order null
point for the poloidal field, which is usually placed at the bottom of
the core plasma, or at the top. The traditional double-null
configuration uses one first-order X-point at the bottom and one at
the top.

In contrast, several advanced divertor configurations have been
proposed where a secondary X-point is included in the divertor
region. These include snowflake-like configurations and the X-point
Target configuration. Snowflake-like configurations approximate a
configuration with an exact second-order null of the poloidal field
dubbed ``snowflake'' \cite{Ryutov2007}. In practice, instead of an
exact second-order null, a configuration is used where two regular
X-points are brought close together, which leads to snowflake-plus and
snowflake-minus configurations \cite{Ryutov2008}. On the other hand,
for an X-Point Target configuration \cite{LaBombard2015}, a secondary
X-point is introduced in the divertor far away from the primary
X-point, in the divertor leg near the target plate.

Each of these divertor configuration is characterized by locations of
the primary and secondary X-points in the domain. The primary X-point
is the most significant as it separates the plasma into the hot core
region and colder scrape-off layer (SOL) region. However, a secondary
X-point helps redirect and distribute the flux of plasma particles and
energy over multiple locations (strike points) on the material
surface.  Furthermore, a secondary X-point may help increase plasma
radiation in the divertor, and potentially it may cause other
interesting and important effects in divertor plasma.


%% file: introduction/grid_generation_for_tokamak.tex
\subsection{Grid generation for tokamak edge plasma transport simulations}

Tokamak boundary and divertor plasma modeling relies heavily on edge
transport modeling codes such as UEDGE \cite{Rognlien1999}, SOLPS
\cite{Wiesen2015}, EDGE2D \cite{Simonini1994}, just to mention a few
major ones. These codes are similar in many ways; they all solve the
time-evolution fluid equations for toroidally-symmetric, collisional
plasma based on the Braginskii equations, using ad-hoc radial
transport coefficients. The simulations are usually carried out in the
actual geometry of a modeled tokamak, to account for details of
magnetic field geometry and plasma-facing components.

Computational grids for tokamak boundary plasma transport modeling are usually
chosen to follow the magnetic flux surfaces, to avoid numerical
pollution caused by the extreme anisotropy of plasma transport along
and across the magnetic field \cite{Umansky2005}. Thus the
computational grid has to follow the underlying magnetic field, and
with one or several X-points present in the simulation domain the grid
topology can become highly nontrivial.

There are several grid generators for tokamak edge plasma region
currently in use. Among those, the UEDGE code uses a grid generator
that is a part of the UEDGE package, SOLPS normally uses grid
generator CARRE \cite{Marchand1996}, and EDGE2D usually relies on grid
generator GRID2D \cite{Taroni1992}. These are sufficient in most cases
for modeling single-null and double-null configurations. However
modeling of advanced divertors may require incorporating secondary
X-points in the divertor region, and grid generators currently in use
for the major edge transport codes are not inherently designed to
produce computational grids for general configurations containing more
than one X-point at arbitrary locations in the domain.

To increment the capabilities for grid generation for tokamak boundary
plasma transport modeling, in particular for advanced divertors, a new grid
generator INGRID has been developed, as described in the present
report.  INGRID (Interactive Grid Generator) is a Python based,
interactive, grid generator for edge plasma transport modeling that is capable
of handling configurations with one or two X-points anywhere in the
computational domain. INGRID provides a robust set of tools such as an
easy to use GUI intended for users of all levels. By internally
handling the challenges that typically arise with generating grids for
tokamak edge plasma region, INGRID can indeed improve efficiency in a
user's workflow for edge-plasma modeling. The INGRID algorithm's
inspiration was drawn from an older IDL-based project Gingred
\cite{Izacard2017} where a ``divide and conquer'', domian partitioning approach for grid
generation was first tried. An important motivating factor for
implementing INGRID in Python was using an open-source language with
advanced numerical and graphical libraries.


%% file: introduction/outline.tex
\subsection{Paper outline}
The aim of this current section is to provide a short introduction to tokamak edge plasma modeling, the state of grid generation for tokamak edge plasma simulations, and motivate the development of INGRID. The next section discusses the classification of equilibria found within a plasma device. Here we define the magnetic topologies that INGRID has been designed to model. Following our discussion on classification, we elaborate on the computational methodology INGRID adopts. Here the interpolation scheme, topology analysis, Patch map construction, and grid construction is detailed. Following our computational algorithm discussion, a benchmark of INGRID is presented. This discussion then leads us into our section on software design. The INGRID package structure, classes, and data formats are discussed. Finally, we summarize our results and point the reader to where INGRID can be obtained.

%% file: classification/all_topology.tex
The geometry of a tokamak (or a similar toroidal plasma device, like a
spheromak or a reversed-field pinch) is defined by the flux function
$\Psi(R,Z)$ such that $\nabla \Psi \cdot \vec{B}
=0$ \cite{Freidberg2014}. Then the poloidal magnetic field components
$B_R,B_Z$ can be expressed as

\beqar
B_R = - \frac{1}{R} \frac{\partial {\Psi}}{\partial Z} \\
B_Z =   \frac{1}{R} \frac{\partial {\Psi}}{\partial R} \\
\eeqar

Surfaces $\Psi(R,Z)$=const form a set of nested magnetic flux surfaces
confining the plasma, and the magnetic field is tangential to the flux
surfaces. The nulls of the poloidal magnetic field, $B_R=B_Z=0$
correspond to extrema or saddle points of the flux function
$\Psi(R,Z)$. The ``magnetic axis'' corresponds to an extremum of
$\Psi$ at the innermost flux surface. Further we consider first-order
saddle points of $\Psi$ where its first derivatives vanish, those
points are ``X-points''.

To understand the range of geometric possibilities in presence of one
and two X-points in the domain, consider the diagrams in
Fig.(\ref{fig:all_conf}).
If there is a single X-point in the region it defines a separatrix
which is a flux surface containing the X-point
Fig. (\ref{fig:all_conf} (a)). The X-point is a self-intersection
point of the separatrix that divides the plane into three
topologically distinct regions. One is the ``core plasma region'',
containing the magnetic axis. Next, there is a region
lying opposite to the core plasma region, across the X-point, and it
is called the ``private flux region'', or PFR. And the remaining part
is the ``common flux region'', or scrape-off-layer (SOL) region.

A second X-point can be added outside of the core plasma region.
Ignoring the degenerate case when the secondary X-point is on the
primary-separatrix, there are two topologically
distinct possibilities: with respect to the primary separatrix the
second X-point can be either in the private flux region,
Fig. (\ref{fig:all_conf} (c)); or in the common flux region
Fig. (\ref{fig:all_conf} (b)). The former case is called
``snowflake-plus'' and the latter case is called ``snowflake-minus''
\cite{Ryutov2007}.

However, from the grid generation perspective, we consider further
variations of ``snowflake-like'' configurations, as shown in
Fig. (\ref{fig:all_conf}). Consider a line orthogonal to flux surfaces
and passing through the secondary X-point $X_2$, and consider the
intersection of this line with the primary separatrix,
$X^{\prime}_2$. For ``snowflake-minus'' configuration there five
possibilities for the projection point $X^{\prime}_2$. It can belong
to: (i) the arc $[M_W,M_E]$ connecting the two midplane-level points,
or (ii) the arc $[M_W,X_1]$, or (iii) the arc $[M_E,X_1]$, or the arc
(iv) $[X_1,S_W]$ connecting the primary X-point and one of the
``strike-points'', or (v) the arc $[X_1,S_E]$. For ``snowflake-plus''
configuration, there are two possibilities for the projection point
$X^{\prime}_2$. It can belong to: (i) $[X_1,S_E]$, or to (ii)
$[X_1,S_W]$.

Based on the location of the secondary X-point $X_2$ and its
orthogonal projection on the primary separatrix $X_2^{\prime}$, we use
the following notation for the configurations with two X-points:

\begin{itemize}
	\item UDN: $SF-$, $X^{\prime}_2 \in [M_E,M_W]$
	\item SF15: $SF-$, $X^{\prime}_2 \in [M_E,X_1]$
	\item SF165: $SF-$, $X^{\prime}_2 \in [M_W,X_1]$
        \item SF45: $SF-$, $X^{\prime}_2 \in [S_E,X_1]$
        \item SF135: $SF-$, $X^{\prime}_2 \in [S_W,X_1]$
        \item SF75: $SF+$, $X^{\prime}_2 \in [S_E,X_1]$
        \item SF105: $SF+$, $X^{\prime}_2 \in [S_W,X_1]$
\end{itemize}

The orthogonal projection and magnetic configurations mentioned above
are seen in
figs. (\ref{fig:all_conf}, \ref{fig:snl_patch_map}-\ref{fig:sf165_patch_map}). The
notation for snowflake-like configurations is inspired by local
geometric analysis of near-snowflake configurations and the numbers
correspond to the geometric angle defining the position of the
secondary X-point \cite{Ryutov2010}. Our geometric classification here
is topology-based, it applies to more general situations when the two
X-points can be far apart; but the notation introduced in
literature\cite{Ryutov2010} is still useful. All in all, with the
single-null (SNL) configuration included, for either one and two
X-points in the domain, there are eight possible configurations. We do
not consider here degenerate cases where the secondary X-point is
exactly on the primary separatrix, or the projected point
$X^{\prime}_2$ is exactly on the primary X-point $X_1$, or it is
exactly on a midplane point $M_E$ or $M_W$; the assumption is that a
practical, experiment-relevant configuration would always have some
finite degree of asymmetry to fall into one of the eight considered
categories.

%% file: computation/divide_conquer.tex
\subsection{Domain partitioning strategy}

The INGRID workflow includes two main steps: (i) constructing a
``skeleton grid'' (also called further a ``Patch-Map'') which
corresponds to the geometry of the magnetic field in hand and consists
of a small number of quadrilateral patches; and (ii) putting a subgrid
on each of the patches. This is illustrated in
Fig. (\ref{fig:patchmap_subgrid}) where a Patch-Map is shown with one
of the patches covered with a subgrid.

The skeleton grid is constructed as a small grid
aligned with the given poloidal magnetic field and respecting the magnetic
field topology. The geometry of magnetic flux surfaces, in particular
the X-points and the magnetic axis, and the geometry of plasma facing
material surfaces all together define a patch map. For calculating a
subgrid, the code divides each quarilateral patch into a number of
radial and poloidal zones according to user-provided input. Finally,
all subgrids are joined together to produce the global grid.

%% file: computation/bicubic_interpolation.tex
\subsection{Magnetic field interpolation}

Input data for grid generation come in the form of the poloidal
magnetic flux $\Psi$ sampled on rectilinear grid in $R,Z$, from an MHD
reconstruction code such as EFIT \cite{Lao_1985} or TEQ
\cite{}. INGRID utilizes a bicubic interpolation implementation
\cite{Press_1992} for obtaining the values of the poloidal magnetic
flux $\Psi$ and its derivatives between data points of a provided
magnetic equilibrium. This functionality is provided by class
\texttt{RectBivariateSpline}
belonging to the \texttt{scipy.interpolate}\cite{virtanen2019scipy,
  scipy.interpolate} package.

Poloidal flux surfaces are reconstructed in INGRID by integrating the ODEs,

\beqar
\dot{R} = -\frac{1}{R} \frac{\partial \Psi}{\partial Z} \\
\dot{Z} = \frac{1}{R} \frac{\partial \Psi}{\partial R}
\eeqar

whereas surfaces orthogonal to poloidal flux surfaces are
reconstructed by integrating the ODES,

\beqar
\dot{R} = \frac{1}{R} \frac{\partial \Psi}{\partial R} \\
\dot{Z} = \frac{1}{R} \frac{\partial \Psi}{\partial Z}
\eeqar

The bicubic interpolation guarantees continuity of first dertivatives
of $\Psi$ at the edges of cells of the original $R,Z$ grid, so the
resulting flux surfaces are smooth.






%% file: computation/refining_reference_points.tex
\subsection{Calculation of reference points}

The topology of magnetic flux surfaces in tokamak edge plasma is
defined by relative positions of X-points and the magnetic axis. These
key reference points are calculated in INGRID by finding nulls of the
poloidal field, solving the equation

\beq
\left( \frac{\partial \Psi}{\partial R} \right)^2 +
\left( \frac{\partial \Psi}{\partial Z} \right)^2 = 0
\eeq

INGRID utilizes method
\texttt{scipy.optimize.root}\cite{virtanen2019scipy,
  scipy.optimize.root_docs} for this purpose. For finding the correct
root of this equation the solver needs an accurate initial guess; the
user is supposed to provide initial guess for the magnetic axis and
for one or two of those X-points that are expected to be included in
the domain.

%% file: computation/topology_analysis.tex
\subsection{Topology analysis}

INGRID performs topology analysis and classification based on the user
specification of one or two X-points in the domain. We first consider
when only one X-point is to be analyzed. The tokamak edge plasma
community often classifies single X-point configurations as ``upper
single null'' or ``lower single null." However, for INGRID there is no
distinction. If one X-point is specified, then INGRID considers the
aforementioned configurations as a general single-null (SNL)
configuration. Instead of using ``lower'' and ``upper'' for the
divertor, and ``inner'' and ``outer'' for target plates, INGRID uses
compass directions North-South-East-West associated with the primary
X-point. The North direction is defined to point into the core plasma
along the $\grad \Psi$ direction, whereas the South direction is
defined to point into the private-flux along the $\grad \Psi$
direction. East and West are then defined orthogonal to the North and
South directions. For example, the inner target plate of an LSN
configuration is in the south-west direction of an SNL configuration,
as illustrated in Fig. (\ref{fig:snl_patch_map}). Similarly, the inner
target plate of a USN configuration is in the south-east
direction. Determining the North-South-East-West directions associated
with the primary X-point is the extent of the analysis performed in
the SNL case.

If two X-points in the domain are to be analyzed, INGRID performs
additional analysis to determine the specific magnetic
topology. INGRID begins by performing the compass analysis from the
SNL case on the primary X-point. Next, INGRID determines whether the
secondary X-point is in the private-flux (PF) region or in the
common-flux region (SOL) with respect to the primary separatrix. If
the secondary X-point resides in the private-flux, the configuration
is of type SF+. If not, the configuration is of type SF-. This
determination is conducted by representing the PF as a generalized
polygon and determining whether a 2D point representing the X-point is
contained within the polygon or not. The vertices of the polygon
correspond to the points obtained from the portion of the limiter
contained between points $S_W$ and $S_E$ (denoted as $[S_W, S_E]$) and
line tracing (detailed in section E) the two arcs $[X_1,S_W]$, $[X_1,
  S_E]$. INGRID implements the point-in-polygon test with the help of
class \texttt{matplotlib.path}. Representing the polygon boundary as a
\texttt{matplotlib.path} object, the method
\texttt{matplotlib.path.contains\_point} can then determine whether
the area enclosed by the path contains the given point or not.  With
the point-in-polygon test completed, an orthogonal projection is
constructed from the secondary X-point to the primary separatrix. This
is done by line tracing equations (3) and (4) where the termination
criteria is intersection with the primary separatrix. Finally, the
specific magnetic topology is determined by the criteria outlined in
Section II for classification of equilibria.



%% file: computation/patch_map_construction.tex
\subsection{Patch-Map construction}

After the analysis of magnetic geometry for a given magnetic field and
establishing what configuration corresponds to it, INGRID creates a
corresponding Patch-Map which is an ordered collection of
quadrilateral patches defining the skeleton grid. A Patch-Map is a 2D
array of patches where one index corresponds to the radial direction,
and the second index corresponds to the poloidal direction; thus the
order of patches in a Patch-Map defines what patches are neighbors of
a given patch in the radial and poloidal directions.

A Patch-Map is constructed using numerical integration and line
tracing of poloidal and radial surfaces through the X-points, and
radial and poloidal surfaces defining the domain boundaries. INGRID
utilizes the method
\texttt{solve\_ivp} from the
\texttt{scipy.integrate}\cite{virtanen2019scipy} package for all
numerical integration purposes. In particular, LSODA\cite{osti_145724}
was selected as the \texttt{solve\_ivp} method of
integration. Equations (1) and (2) provide INGRID the capability to
reconstruct poloidal flux, whereas equations (3) and (4) provide
INGRID the capability to reconstruct radial flux surfaces. The process
of flux surface reconstruction and, when applicable, visualization of
the reconstruction we refer to as line tracing. Radial domain
boundaries are flux surfaces corresponding to maximum and minimum
values of the poloidal flux function $\Psi$, these values are defined
by the user in the input file. The poloidal domain boundaries are
target plates surfaces, also defined in the input file. 
For each of the divertor configurations that INGRID can use - which
includes a single-null, unbalanced double-null, and six snowflake-like
configurations - there is a specific type of Patch-Map that defines
the topology of this configuration. For example, for a SNL
configuration shown in Fig. (\ref{fig:snl_patch_map}), a Patch-Map
includes two radial zones, two poloidal zones defining divertor legs,
and four poloidal zones defining the edge plasma domain around the
last closed flux surface. Such a Patch-Map that contains twelve
patches is sufficient to represent a general single-null geometry,
albeit in a most basic and crude way. Note that one could use only ten
patches to represent the single-null topology, but using twelve
patches allows matching more accurately the general shape of a
single-null configuration. Furthermore, a finer grid representing a
single-null geometry can be always represented as this Patch-Map with
local refinement applied to one or several of these twelve patches.
For a more complicated unbalanced double null (UDN) geometry shown in
Fig. (\ref{fig:udn_patch_map}), the Patch-Map must include three
radial zones because there are two separatrices there. There are four
poloidal zones to represent four divertor legs there, and together
with four poloidal zones covering the core domain there are total
eight poloidal zones. All in all, there are twenty four patches in
this case. Similarly, for each of the snowflake-like configurations
there are twenty four patches, as illustrated in
Figs. (\ref{fig:sf15_patch_map}-\ref{fig:sf165_patch_map}).
It is important to comment on the consistency of boundaries of each Patch with adjacent Patches. Regardless of divertor configuration, each Patch map generation method enforces that boundaries align radially by simply defining adjacent Patch boundaries in terms of the same boundary. Poloidal consistency of Patch boundaries along the north and south is attained by simply continuing line tracing from the end point of the adjacent Patch. It should be noted that this process cannot guarantee the consistency of Patch boundaries along the upper core. This boundary mismatch, however, is directly related to line tracing tolerance and can be controlled by the \texttt{tol} attribute found within the integrator settings of the parameter file. 

%% file: computation/grid_construction.tex
\subsection{Grid construction}

For a given magnetic configuration, a Patch-Map represents a very
crude grid where each Patch is a ``quadrilateral'' with four vertices.
The radial sides of this quadrilateral are defined by two flux
surfaces, $\Psi(R,Z)=\Psi_1$ and $\Psi(R,Z)=\Psi_2$. The poloidal
sides of a Patch are usually constructed to be aligned with $\nabla
\Psi$, which makes the skeleton grid locally orthogonal. 

However, more generally the poloidal sides of a Patch can deviate from
the $\nabla \Psi$ direction. For example, for those patches that
contain the poloidal boundaries of the domain, given by the target
plates, one of the sides is defined by the target plate shape. The
curve describing the target plate can be arbitrary, as long as it does
not form ``shadow regions'', i.e., $\Psi$ is a monotonic function of
the length along the plate.

Going beyond the skeleton grid, a Patch can be divided in a number of
radial and poloidal zones, forming a subgrid local to this patch. The
radial zones are constructed to be aligned with flux surfaces, so the
global grid remains aligned with the poloidal magnetic field. For the
poloidal zones, the main algorithm is based on dividing the patch
poloidally into uniform length line segments. However, as described further,
there are options in the code for controlling the radial and poloidal
distribution of subgrid.

The radial and poloidal dimensions and distribution of subgrid on a
given Patch are not entirely independent of subgrids on other patches
as the global grid still has to be Cartesian in the index space. Thus
the poloidal grid has to be consistent for those patches that are
stacked on top of each other radially, and the radial grids have to be
consistent for those patches stacked on top of each other poloidally.

%% file: computation/grid_optimization.tex
\subsection{Grid customization}

INGRID provides users a number of tools for customization of generated
grids, which can be controlled via the parameter file. Users can
modify both the Patch-Map and the subgrids in order to optimize the
global grid.

Default settings for constructing a Patch-Map use the horizontal plane
through the magnetic axis, commonly referred to as the midplane.  For
grid generation purposes, INGRID allows shifting the effective
magnetic axis location vertically and horizontally. Moreover, instead
of using the horizontal directions for the midplane, the user can set
two angles defining a ``generalized midplane''. Since the midplane can
be thought of as two rays emanating from the magnetic axis with angles
$0$ and $\pi$ radians, the ``generalized midplane" is defined
similarly but with user-defined ``effective magnetic axis'' location,
and with user-specified angles for both rays.

Also, for those patches that include an X-point as one of their
vertices, the default poloidal boundaries use in the East and West
directions from the X-point along the $\grad \Psi$ direction. However,
the user can redefine those curved, replacing them by straight lines
in a desired direction.

For fine-tuning the grid, subgrids can be adjusted as well. By
default, during Patch refinement, grid seed-points are distributed
uniformly in length along the radial boundaries and
distributed uniformly along the poloidal boundaries in
locally-normalized $\Psi$. This default behavior can be changed so
that grid seed-point placement obeys a user specified distribution
function.

Another important customization feature in INGRID is a
``\texttt{distortion\_correction}" tool for mitigating grid
shearing. This tool allows the user to bound the angles found within
the interior of a grid cell in order to generate nearly orthogonal
subgrids in patches. The implementation is as follows. Let a
quadrilateral cell of a subgrid be defined by four nodes $A, B, C,$
and $D$ where $AB$ is the top-face, $BC$ is the right-face, $CD$ is
the bottom-face, and $DA$ is the left-face of the cell. The angle
$\measuredangle DAB := \alpha$ is measured. If $\alpha$ is not within
a user-defined range $[\alpha_{\min}, \alpha_{\max}]$, then the node
$D$ is translated along the poloidal flux surface until $\alpha$ is
within the range. The direction of translation is determined by
whether $\alpha \geq \alpha_{\max}$ or $\alpha \leq
\alpha_{\min}$. During the translation of $D$, we must also avoid
collision with a neighboring node on the poloidal flux surface. To
ensure collision does not occur, the translation stops when the moving
node $D$ comes within $\epsilon$ of a neighboring node. Upon
termination, the current value of $\alpha$ is used to define the
transformed cell. An example of the \texttt{distortion\_correction}
feature applied to a grid can be seen in Fig.(
\ref{fig:distortion_correction}).

%% file: performance/scaling_time.tex
\subsection{Scaling of calculation time}
The results of INGRID timing test are shown in
Table (\ref{tab:benchmark_table_1}) and in
Fig. (\ref{fig:benchmark_grid_scaling}). 
These tests were run on an Apple Mac Mini (2020 model) housing an Apple M1 chip (3.2 GHz processor).
The scaling appears sublinear
for small grids; for larger grids it asymptotes to linear. Note that
the cost of grid generation is not significant in a typical edge
plasma modeling workflow; running the simulation takes orders of
magnitude more computing time. This scaling makes INGRID practical for constructing large grids.\\

\begin{table}[H]
\begin{tabular}{|c|c|c|}
    \toprule
    \multicolumn{3}{|c|}{SNL}\\
    \cline{1-3}
    {Cells Per Patch} & Total Cells &    Time (s) \\
    \midrule
    9  &       108 &   8.02 \\
    16  &       192 &   11.44 \\
    25  &       300 &   14.79 \\
    36  &       432 &  18.29 \\
    49  &       588 &  21.58 \\
    64  &       768 &  25.02 \\
    81  &       972 &   28.49 \\
    100  &      1200 &  32.27 \\
    121  &      1452 &  35.32 \\
    144  &      1728 &  38.61 \\
    169 &      2028 &  41.98 \\
    196 &      2352 &  45.46 \\
    225 &      2700 &  49.08\\
    \bottomrule
\end{tabular}
\begin{tabular}{|c|c|c|}
    \toprule
    \multicolumn{3}{|c|}{SF75}\\
    \cline{1-3}
    {Cells Per Patch} & Total Cells &   Time (s) \\
    \midrule
    9  &       243 &   21.32 \\
    16  &       432 &   27.69 \\
    25  &       675 &  34.45 \\
    36  &       972 &  41.10 \\
    49  &      1323 &  48.31 \\
    64  &      1728 &  54.82 \\
    81  &      2187 &  62.24 \\
    100  &      2700 &  68.69 \\
    121  &      3267 &  75.44 \\
    144  &      3888 &  82.63 \\
    169 &      4563 &  89.97\\
    196 &      5292 &  96.21\\
    225 &      6075 &  103.71\\
    \bottomrule
\end{tabular}
\caption{A benchmark of grid generation for both an SNL and SF75 configuration. Grids were generated with $n\times n$ many cells per Patch with $n = \{3, 4, 5, \dots, 15\}$. With $n \times n$ subgrid dimensions, SNL configurations contain $12n^2$ many cells, whereas SF75 configurations contain $27n^2$ many cells.}
\label{tab:benchmark_table_1}
\end{table}

%% file: performance/benchmarks.tex
\subsection{Benchmark testing}

To verify grids calculated by INGRID, several benchmark tests have
been performed with the UEDGE code.

In these testes, UEDGE solutions were compared, using grids from
INGRID and grids produced with the UEDGE internal grid generator; for
the same physics problem statement, the same boundary conditions,
using the same (or very close) domain geometry.

In one of these test problems, a snowflake-like SF75 configuration was
used, based on magnetic reconstruction data from the TCV tokamak. The
UEDGE grid generator cannot deal with the SF75 configuration directly;
but it can be set up to treat each X-point as a part of a separate SN
configuration, and then joining two such SN grids together one can
produce a grid for the full SF75 domain.

The UEDGE code was set up to solve to the steady state the
time-evolution equations for plasma density, plasma parallel momentum,
ion thermal energy, and electron thermal energy.

\beq
\pdiff{}{t} n_i + \grad \cdot 
\left[ n_i \vec{V}_i \right]
= S_i
\eeq

\beq
\pdiff{}{t} \left[ M n_i {V}_{i,||} \right] + \grad \cdot
\left[
M n_i \vec{V}_i {V}_{i,||}  - \hat{\eta}_i \grad {V}_{i,||}
\right]
= S_{m,||}
\eeq

\beq
\pdiff{}{t}
\left[
\frac{3}{2} n T_i
\right]
+
\grad \cdot
\left[
\frac{5}{2} n_i T_i \vec{V}_i
+
\vec{q}_i
\right] = S_{E,i}
\eeq

\beq
\pdiff{}{t}
\left[
\frac{3}{2} n T_e
\right]
+
\grad \cdot
\left[
\frac{5}{2} n_e T_e \vec{V}_e
+
\vec{q}_e
\right] = S_{E,e}
\eeq

Here we use the standard notation: $n_i$ is the plasma density,
$\vec{V}_i$ is the plasma fluid velocity, $T_{e,i}$ is the electron
and ion temperature, $\vec{q}_{e,i}$ is the electron and ion heat
flux, $S_i$, $S_{m,||}$, $S_{E,e,i}$ are sources of plasma density,
parallel momentum, and electron and ion thermal energy
\cite{Rognlien1999}.

The grids used for the calculation are shown in
Fig. (\ref{fig:ingrid_grid}). The grid generated with the UEDGE grid
generator is constructed to be strictly locally orthogonal; the grid
from INGRID is not orthogonal. Also, there is slight difference in the
domain shape; the grid from INGRID uses flat target plates while the
grid from UEDGE uses curved target plates orthogonal to $\Psi$. Still,
the steady state solutions exhibit essentially the same distributions
of plasma density, temperature, and parallel flow velocity, as can be
seen in Fig. (\ref{fig:benchmark_collection}). Beyond that, a
quantitative comparison of radial plasma profiles at the outer
midplane was carried out, with reassuring results indicating abscence
of any major issue in INGRID grids.

%% file: software_methodology/software_organization.tex
\subsection{\label{sec:label2}INGRID package}
INGRID has been exclusively developed in the Python programming
language to take advantage of the free, community supported graphical and numerical libraries, and due to the increasing popularity in major tokamak plasma modeling projects such as OMFIT\cite{Meneghini_2015,
  Orso_MENEGHINI2013} and PyUEDGE \cite{PyUEDGE}.
The \texttt{Ingrid} class contained within the
\texttt{ingrid} module provides the primary API for users. This \texttt{Ingrid} class is used to activate
INGRID's GUI mode and also contains high-level methods for importing
data, visualizing data, analyzing data, grid-generation, and exporting
of data; all of which can be utilized noninteractively in Python scripts. Class
\texttt{IngridUtils} is contained within the \texttt{utils} module and
serves as the base class for \texttt{Ingrid}. \texttt{IngridUtils}
class methods encapsulate much of the lower-level software details
used to implement the methods in the \texttt{Ingrid} class. Because of
this, \texttt{IngridUtils} is encouraged for use by advanced users and
developers of INGRID. In addition to \texttt{IngridUtils}, class
\texttt{TopologyUtils} can be found within the \texttt{utils}
module. In a manner similar to \texttt{IngridUtils}, the
\texttt{TopologyUtils} class serves as a base class for each
magnetic-topology class within the \texttt{topologies}
subpackage. \texttt{TopologyUtils} contains key methods for generating
Patch-Maps, visualizing data, generating grids, and exporting grids in
gridue format. Eight magnetic-topology classes are contained within
their own modules within the \texttt{topologies} subpackage:
\texttt{SNL}, \texttt{UDN}, \texttt{SF15}, \texttt{SF45},
\texttt{SF75}, \texttt{SF105}, \texttt{SF135}, and
\texttt{SF165}. Each magnetic-topology class contains configuration
specific line-tracing instructions for construction of Patch-Maps,
Patch-Map layout information, and gridue formatting
information. \texttt{Ingrid} and \texttt{IngridUtils} conduct analysis
of MHD equilibrium data in order to decide which magnetic-topology
class to instantiate from the \texttt{topologies} subpackage. The
\texttt{IngridUtils} class always maintains a reference to the
instantiated object in order to effectively manage
grid-generation.\\ \indent All GUI operation is managed by class
\texttt{ingrid\_gui} within the \texttt{gui}s subpackage. INGRID's GUI
front-end was developed with the Tkinter package; a Python interface
to the Tk GUI toolkit that is available within the Python Standard
Library. Class \texttt{ingrid\_gui} is simply responsible for managing
event handling, and managing an \texttt{Ingrid} object that is used to
drive the GUI with direct calls to the available high-level
methods.\\ \indent Beyond modules \texttt{ingrid} and \texttt{utils},
modules \texttt{geometry}, \texttt{interpol}, and
\texttt{line\_tracing} form the computation and modeling foundation of
INGRID. Class \texttt{EfitData} can be found
within the \texttt{interpol} module. Class \texttt{EfitData}
is used to provide an interpolated representation of provided MHD
equilibrium data. \texttt{EfitData} computes partial derivative
information of MHD equilibrium data, provides interpolated $\Psi$
function values by interfacing with class \texttt{scipy.interpolate.RectBivariateSpline}, and
contains methods for visualisation of interpolated MHD equilibrium
data. Module \texttt{geometry} contains classes \texttt{Point},
\texttt{Line}, \texttt{Patch}, and \texttt{Cell}. These classes are
the building-blocks for creation of Patch-Maps and generation of
grids.

%% file: software_methodology/geometry_hierarchy.tex
\subsection{\label{sec:label2}INGRID geometry object hierarchy}
If we are to adopt an object-oriented approach to grid generation,
then we must develop a set of tools that can be utilized throughout
our project. Here we discuss how INGRID defines a collection of
geometric classes in order to make the grid generating process as
simple as possible for all magnetic topologies of interest. To do so,
INGRID defines the following collection of geometric abstractions: the
Point class, the Line class, the Cell class, and the Patch class. All
together, these classes arm INGRID with the ability to represent any
magnetic topology of interest. We provide a very brief description of
the classes here. Figure \ref{fig:patchmap_subgrid} illustrates the
geometry collection described below.
A Point object simply represents an arbitrary $(R,Z)$ spatial
coordinate in the computational domain. Along with $(R,Z)$, coordinates, $\Psi$ values can be returned from the Point object. 
A Line object object is defined by a list of two or more Point
objects. This Line object definition allows for the representation of
any arbitrary curve we may encounter (e.g. constant $\Psi$ surface,
target plate, limiter geometry). This can be done since the collection of Point objects correspond to segments that are the discretization of the curve of interest.
A Patch object represents a closed region of the domain, and a block of the block-structured mesh INGRID computes. Patches are defined by four Line
objects. The ``North" Line (top), ``East" Line (right), ``South" Line (bottom), and ``West" Line (left) define the Patch borders which form a closed, clockwise-oriented loop. Methods of the Patch class assist with visualization and computation. For example, Patch method \texttt{make\_subgrid} directly handles grid generation for the local region of interest. 
A Cell object resides within a Patch and represents a quadrilateral grid cell. Cells are defined by five Point objects:
four corners (NW, NE, SW, SE) and a center. These Cell objects provide the spatial and
experimental data that are written to exported grid files.

From these definitions, we have the building blocks for modeling any
of the magnetic topologies of interest we mentioned in the previous
section. In particular, we aim to construct a collection of Patch
objects representing the divertor configuration of interest. We call
this collection of Patch objects a Patch-Map. This Patch-Map allows us
to create a grid of Cell objects within each Patch, thus providing the
final grid. The management of the Patch-Map creation and grid
generation is managed by a magnetic topology class of modeling
interest. As of the current INGRID release, we have defined magnetic topology
classes SNL, UDN, SF15, SF45, SF75, SF105, SF135, and SF165. These are
contained within a dedicated topologies subpackage within the INGRID
code.

%% file: software_methodology/ingrid_parameter_file.tex
\subsection{INGRID parameter file}
We have decided to use YAML formatted files for the parameter
file\cite{PyYAML} for user control INGRID in GUI mode and restoring
sessions in non-interactive scripting mode. This YAML file is similar
to the familiar Fortran namelist files due to the key-value structure
it employs. YAML is in an easy to read format that has extensive
support within Python. With the PyYAML library, Python reads a YAML
formatted file and internally represents it as a Python
dictionary. This allows users to model cases in the INGRID GUI and
reuse a parameter file in scripting for later usage (e.g. batch grid
generation). Some key controls within the parameter file include: EFIT
file specification, specification of number of X-points, approximate
coordinates of X-point(s) of interest, approximate magnetic-axis
coordinates, $\Psi$ level values, and target plate settings (files,
transformations). Other controls in the parameter file include: path
specification for data files, grid cell np/nr values, poloidal and
radial grid transformation settings, limiter specific settings,
saving/loading of Patch maps, gridue settings, and debug
settings. This is not an exhaustive list. Further details can be found
in INGRID's Read The Docs online documentation.

%% file: software_methodology/ingrid_target_plate_file.tex
\subsection{INGRID target plate file}

INGRID users must specify the geometry of the limiter and/or target
plates (by which we mean here the entire first-wall contour
surrounding the plasma), to represent the shape of material walls in a
modeled device. The limiter and target plates are represented in
INGRID by a piecewise-linear model defined by a set of nodes; the
(R,Z) coordinates of those nodes are expected to be provided in
separate data files. There is one data file for the limiter and one
for each target plate, either in the text format or as a NumPy
binary. The names of those data files are set in the INGRID parameter
file. In the case that the limiter and target plate data are provided
in text format, the user must specify the (R,Z) coordinates for each
point defining the surface sequentially on a separate line in the
corresponding data file; and Python-formatted single- line comments
can be included, as shown in the Appendix.

For use of NumPy binary files, users must also adhere to a particular
internal file structure. Given two NumPy arrays of shape $(n, )$ that
represent $R$ and $Z$ coordinate values respectively, one can define a
NumPy array of shape $(2,n)$ representing the $n$-many points required
to model the piecewise-linear model of interest. This NumPy array of
shape $(2, n)$ can be saved into a NumPy binary file in order to be
loaded into INGRID.  In addition to the requirements above, INGRID
asserts that strike-point geometry files used for Patch-Maps are
monotonic in $\Psi$ along the length of the target plates (i.e. no
shadow regions). The requirement of target plates to be monotonic in
$\Psi$ allows INGRID to parameterize the $(r, z)$ coordinates of the
geometry with the $\Psi$ values. With a unique mapping of $\Psi$ to
$(r, z)$ target plate coordinates, INGRID can generate Patch objects
that conform exactly to the plate or limiter geometry.  While
operating INGRID in GUI mode, users will be warned if the loaded
geometry file is not monotonic in $\Psi$ along the target plate
length.

%% file: software_methodology/ingrid_workflow.tex
\subsection{INGRID workflow}
INGRID can be operated via GUI or utilizing the INGRID library
directly in Python scripts. The GUI workflow highlights the
interactive nature of INGRID by allowing users to visually inspect MHD
equilibrium data, configure geometry, and refine parameter file values
on the fly. Figure \ref{fig:ingrid_gui} shows the simple GUI with both
MHD equilibrium data, and the corresponding grid plotted (text editor
not pictured, see Appendix for parameter file). For both GUI operation
and scripting with INGRID, the high-level INGRID workflow is: (i)
Parameter file visualization and editing, (ii) Analysis of MHD
equilibrium data and creation of Patch-Map, (iii) Patch-Map refinement
and gridue export.

INGRID internally handles step (ii) and leaves the user to with steps
(i) and (iii). These steps are where the user is able to customize the
Patch-Map and grid to meet their modeling needs.

Step (i) in the INGRID workflow allows users to visually inspect MHD
equilibrium data, target-plates and limiter geometry, and $\Psi$-level
contours that are specified within a loaded parameter file. Since
creation of grids is tied directly to MHD equilibrium analysis and
Patch-Map creation, step (i) is crucial for successful grid
generation. To simplify this step, the INGRID GUI provides an easy to
use environment for preparation of a parameter file for the subsequent
analysis of MHD equilibrium data and Patch-Map creation. Examples of
common operations at this step include modifications to strike-point
geometry and $\Psi$-level boundaries for subsequent Patch-Maps. Once a
user is satisfied with parameter file settings, step (ii) can be
immediately executed with no further user intervention. Should any
errors in Patch-Map creation occur (e.g. misplaced target-plates,
$\Psi$-boundaries that do not conform to configuration specific
requirements), INGRID will prompt the user and allow for appropriate
edits to be made. Upon completion of step (ii), the created Patch-Map
will be provided to users as a new matplotlib figure. From here the
user can decide to proceed with Patch-Map refinement or start over at
step (i) to make edits to the Patch-Map.

In order to streamline grid generation and skip directly to step
(iii), INGRID supports Patch-Map reconstruction. This feature allows
users to bypass line-tracing routines by reloading a saved Patch-Map
from a previous INGRID session. To do so, INGRID encodes essential
geometry and topology analysis data in a specially formatted
dictionary that is then saved as a NumPy binary file. Class
\texttt{IngridUtils} handles the encoding and reconstruction of
Patch-Maps. These reconstruction features can be configured by the
user within the INGRID parameter file.

After a Patch-Map has been generated or reconstructed, users can
configure grid generation specific settings that will be utilized
during Patch-Map refinement. Similar to Patch-Map generation, once all
local subgrids have been created within Patch objects, a new
matplotlib figure is presented with the generated grid. From here,
users can make grid generation setting edits in the INGRID parameter
file or proceed to exporting a gridue file.


%% file: closing/conclusion.tex
INGRID is a new grid generator for tokamak boundary region, it is
capable of producing grids for single-null (SNL), unbalanced
double-null (UDN), and snowflake-like (SF) configurations. Currently,
exported grids are in the format of the UEDGE code, as detailed in
Ref. (\cite{Rensink2017}); future development may include addition of
grid formats used by other codes if INGRID is adopted in the broader
edge-plasma community, beyond UEDGE. INGRID can be utilized via the
INGRID Python package, or through a parameter file driven GUI
mode. Source code as well as documentation is publicly available on
GitHub\cite{ingrid_github} and Read the Docs\cite{ingrid_readthedocs}
respectively. The internal equilibrium topology analysis algorithm
provides
the ability to automatically identify the divertor configuration
embedded within experimental data with minimal user interaction. The 
geometry class hierarchy approach to domain partitioning and Patch-Map
abstraction is an essential component of INGRID and results in a modular
approach to grid generation. These localized grids are combined into a
global grid that are then ready for export. Current computational
scaling of grid generation algorithm follows a sublinear trend
independent of magnetic-topology modeled. Benchmarking of INGRID
against the internal grid generator in UEDGE is demonstrated for an
SF75 snowflake-like configuration. These tests illustrate INGRID's
ability to produce practical grids for tokamak edge modeling, for
complex magnetic flux function with one or two X-points in the domain,
and for nontrivial target plate geometry.

\section{Acknowledgments}
The authors would like to thank M.E.Rensink for his help with grid
generation in UEDGE, and L.L.LoDestro for many critical comments on
the manuscript. This work was performed for U.S. Department of Energy
by Lawrence Livermore National Laboratory under Contract
DE-AC52-07NA27344, and General Atomics under Contract
DE-FG02-95ER54309.

%% file: appendix.tex
\onecolumngrid
An example INGRID parameter file is shown below.
In the file, the user is supposed to provide settings relevant to grid and Patch generation:
1) For the magnetic field geometry, the name (with path) of data file, in the commonly used neqdsk format;
2) For radial boundaries, the values of the normalized poloidal flux $\Psi$ for each of the radial boundaries;
3) For poloidal boundaries, the code has options to use one of these: 
	a) limiter data embedded in the eqdsk file
	b) limiter data provided in a separate file (the file name and path must be included)
	c) target plates geometry in separate files, one per each plate (the file names and paths must be included);
4) How many X-points to include in the domain, 1 or 2;
5) Approximate R,Z coordinates for each of the included X-points and for the magnetic axis, to provide initial guess for the solver;
6) Dimensions of sub-grids;
7) Options related to grid customization;
8) Options related to integrator settings.

\begin{lstlisting}[basicstyle=\small, aboveskip=\bigskipamount, frame=single, captionpos=b, caption={The YAML formatted configuration file. YAML files utilize Python formatted comments, keyword-value mappings, and nesting of structures via indentation.}]
# ------------------------------------------------------------------
# User data directories
# ------------------------------------------------------------------
dir_settings:
  eqdsk: ../data/SNL/DIII-D/  # dir containing eqdsk
  limiter: .  # dir containing limiter
  patch_data: ../data/SNL/DIII-D/  # dir containing patch data
  target_plates: ../data/SNL/DIII-D/ # dir containing target plates
# ------------------------------------------------------------------
# eqdsk file name
# ------------------------------------------------------------------
eqdsk: neqdsk
# ------------------------------------------------------------------
# General grid settings
# ------------------------------------------------------------------
grid_settings:
  # ----------------------------------------------------------------
  # Settings for grid generation 
  # (num cells, transforms, distortion_correction)
  # ----------------------------------------------------------------
  grid_generation:
    distortion_correction:
      all:
        active: True # true, 1 also valid.
        resolution: 1000
        theta_max: 120.0
        theta_min: 80.0
    np_default: 3
    nr_default: 3
    poloidal_f_default: x, x
    radial_f_default: x, x
  # ---------------------------------------------------------------
  # guard cell size
  # ---------------------------------------------------------------
  guard_cell_eps: 0.001
  # ---------------------------------------------------------------
  # num levels in efit plot
  # ---------------------------------------------------------------
  nlevs: 30
  # ---------------------------------------------------------------
  # num xpts
  # ---------------------------------------------------------------
  num_xpt: 1
  patch_generation:
    strike_pt_loc: target_plates # 'limiter' or 'target_plates'
    rmagx_shift: 0.0
    zmagx_shift: 0.0
  # ---------------------------------------------------------------
  # Psi levels
  # ---------------------------------------------------------------
  psi_1: 1.066
  psi_core: 0.95
  psi_pf_1: 0.975
  # ---------------------------------------------------------------
  # magx coordinates
  # ---------------------------------------------------------------
  rmagx: 1.75785604
  zmagx: -0.0292478683
  # ---------------------------------------------------------------
  # xpt coordinates
  # ---------------------------------------------------------------
  rxpt: 1.300094032687
  zxpt: -1.133159375302
  # ---------------------------------------------------------------
  # Filled contours vs contour lines
  # ---------------------------------------------------------------
  view_mode: filled
# -----------------------------------------------------------------
# Saved patch settings
# -----------------------------------------------------------------
patch_data:
  file: LSN_patches_1597099640.npy
  preferences:
    new_file: true
    new_fname: LSN_patches_1597099640.npy
  use_file: false
# -----------------------------------------------------------------
# Integrator
# -----------------------------------------------------------------
integrator_settings:
  dt: 0.01
  eps: 5.0e-06
  first_step: 5.0e-05
  max_step: 0.064
  step_ratio: 0.02
  tol: 0.005
# -----------------------------------------------------------------
# Limiter settings
# -----------------------------------------------------------------
limiter:
  file: ''
  use_efit_bounds: false
# -----------------------------------------------------------------
# target plate settings
# -----------------------------------------------------------------
target_plates:
  plate_E1:
    file: d3d_otp.txt
    zshift: -1.6
  plate_W1:
    file: d3d_itp.txt
    zshift: -1.6
\end{lstlisting}

%% file: figures.tex
\newpage

\begin{figure}[H]
    \centering
    \includegraphics[width=\linewidth]{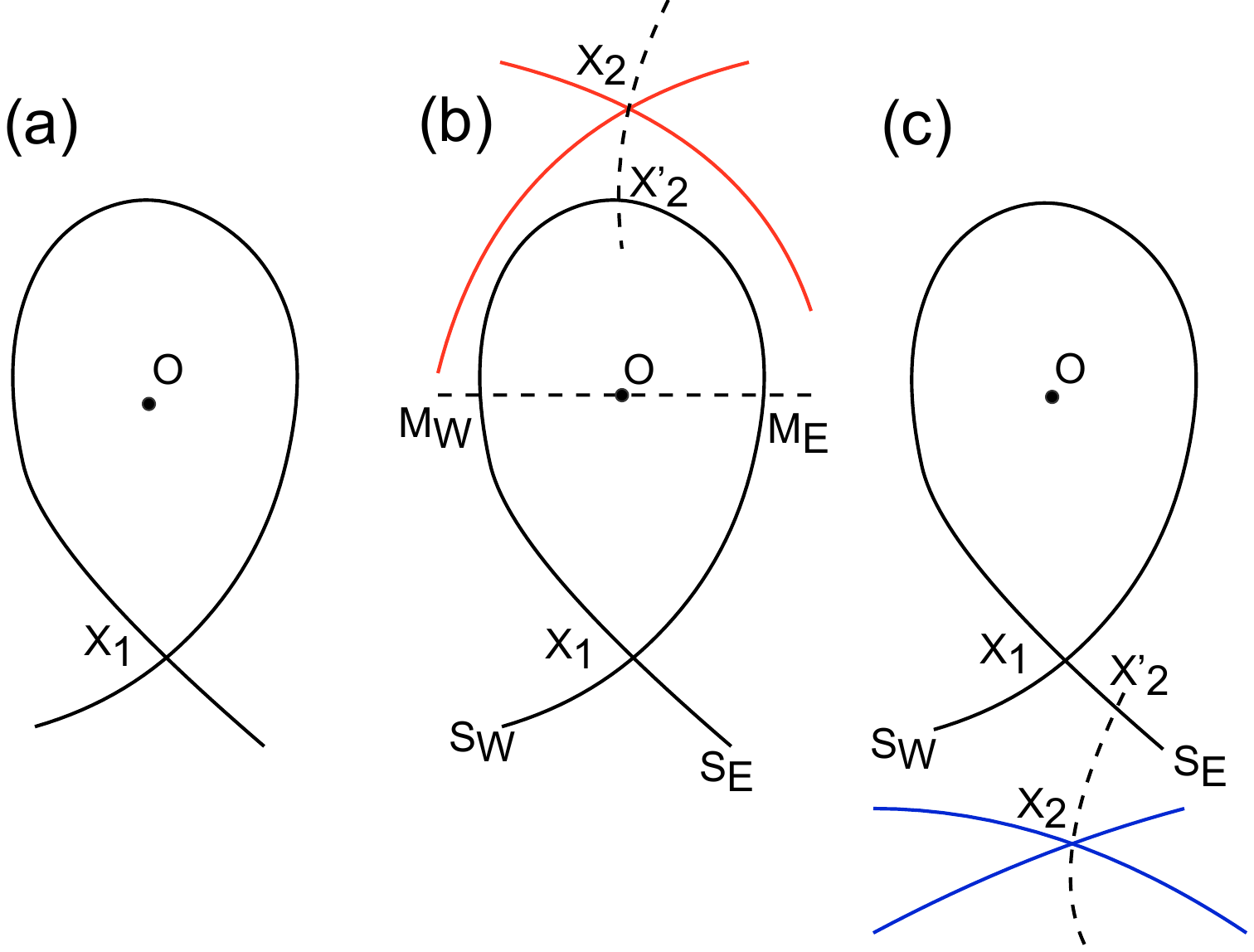}

    \caption{Topological possibilities with one and two X-points. As
    explained in the main text, case (a) is a single-null
    configuration; case (b) with the secondary X-point in the
    common-flux region has five variations, depending on the location
    of the projection point $X_2^{\prime}$; case (c) with the
    secondary X-point in the private-flux region has two variations
    depending on the location of the projection point $X_2^{\prime}$.
    The East-West notation for the two midplane points and the two
    strike points is based on designating the direction from the
    primary X-point toward the O-point as ``North''. This is invariant
    notation, independent on whether the primary X-point is at the
    top, at the bottom, or anywhere else.}

    \label{fig:all_conf}
\end{figure}

\begin{figure}[H]
    \centering
        \includegraphics[width=\textwidth]{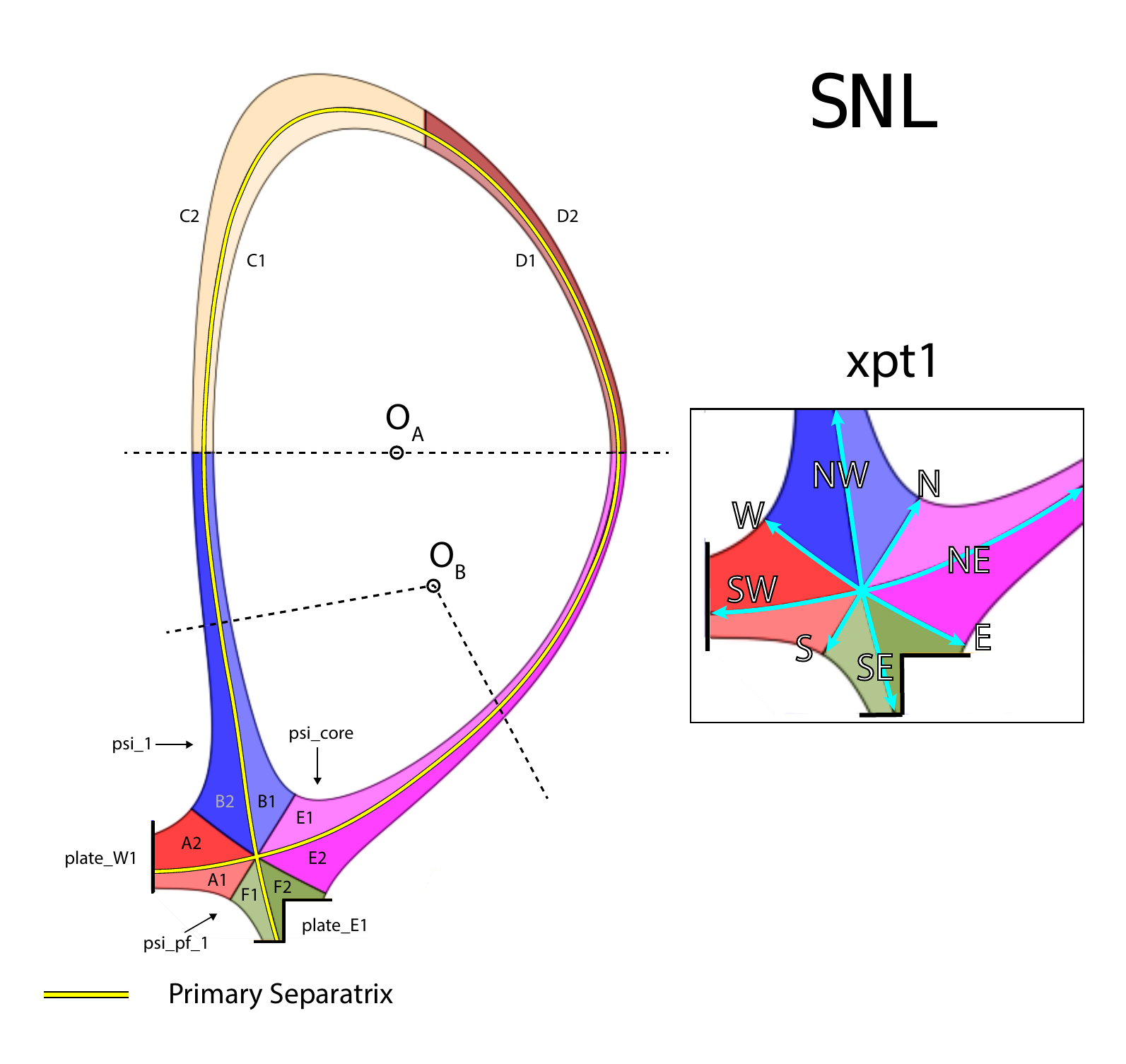}
        \caption{SNL Patch-Map with the generalized midplane overlayed. $O_A$ shows the 
        original magnetic axis used to define the horizontal midplane through $O_A$. The midplane
        defines the poloidal boundary between patches B/C and D/E. For Patch map customizability, INGRID allows for a generalized midplane through $O_B$ to be defined (described in Section II-G). This is done by redefining the location of the magnetic axis and the directions of the rays generating the midplane.}
        \label{fig:snl_patch_map}
\end{figure}

\begin{figure}[H]
    \centering
        \includegraphics[width=\textwidth]{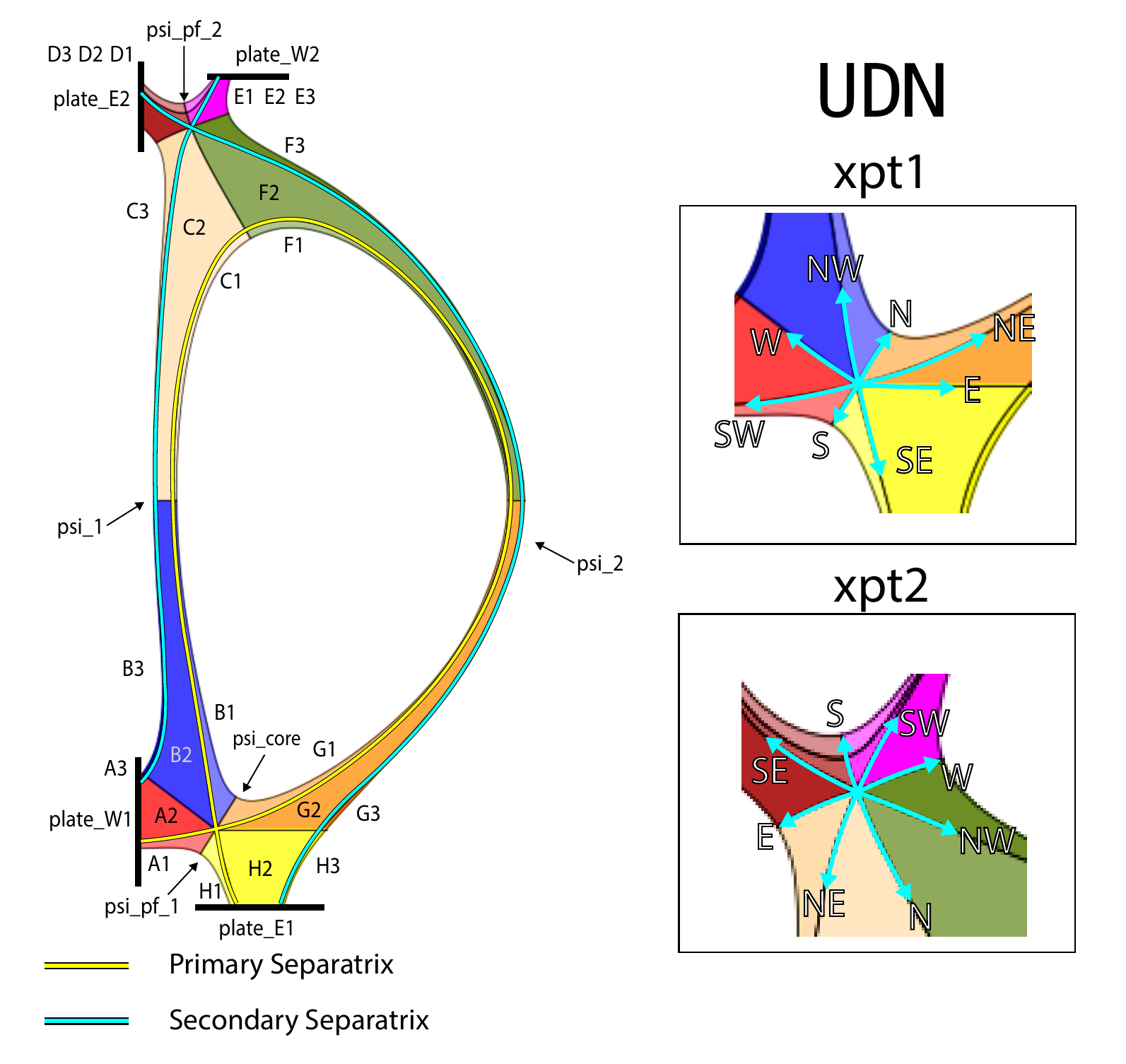}
        \caption{UDN Patch-Map}
        \label{fig:udn_patch_map}
\end{figure}
\begin{figure}[H]
    \centering
        \includegraphics[width=\textwidth]{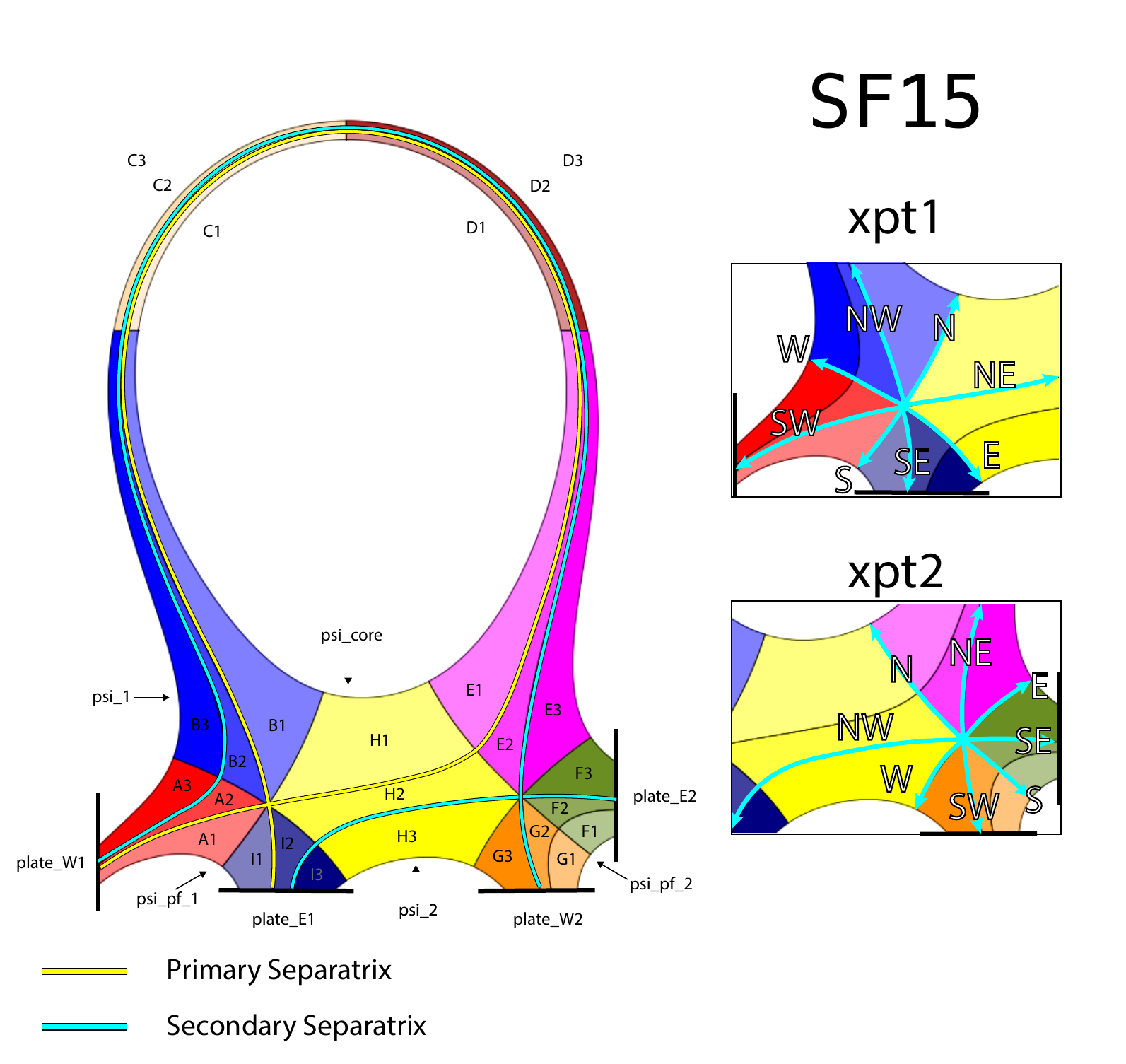}
        \caption{SF15 Patch-Map}
        \label{fig:sf15_patch_map}
\end{figure}
\begin{figure}[H]
    \centering
        \includegraphics[width=\textwidth]{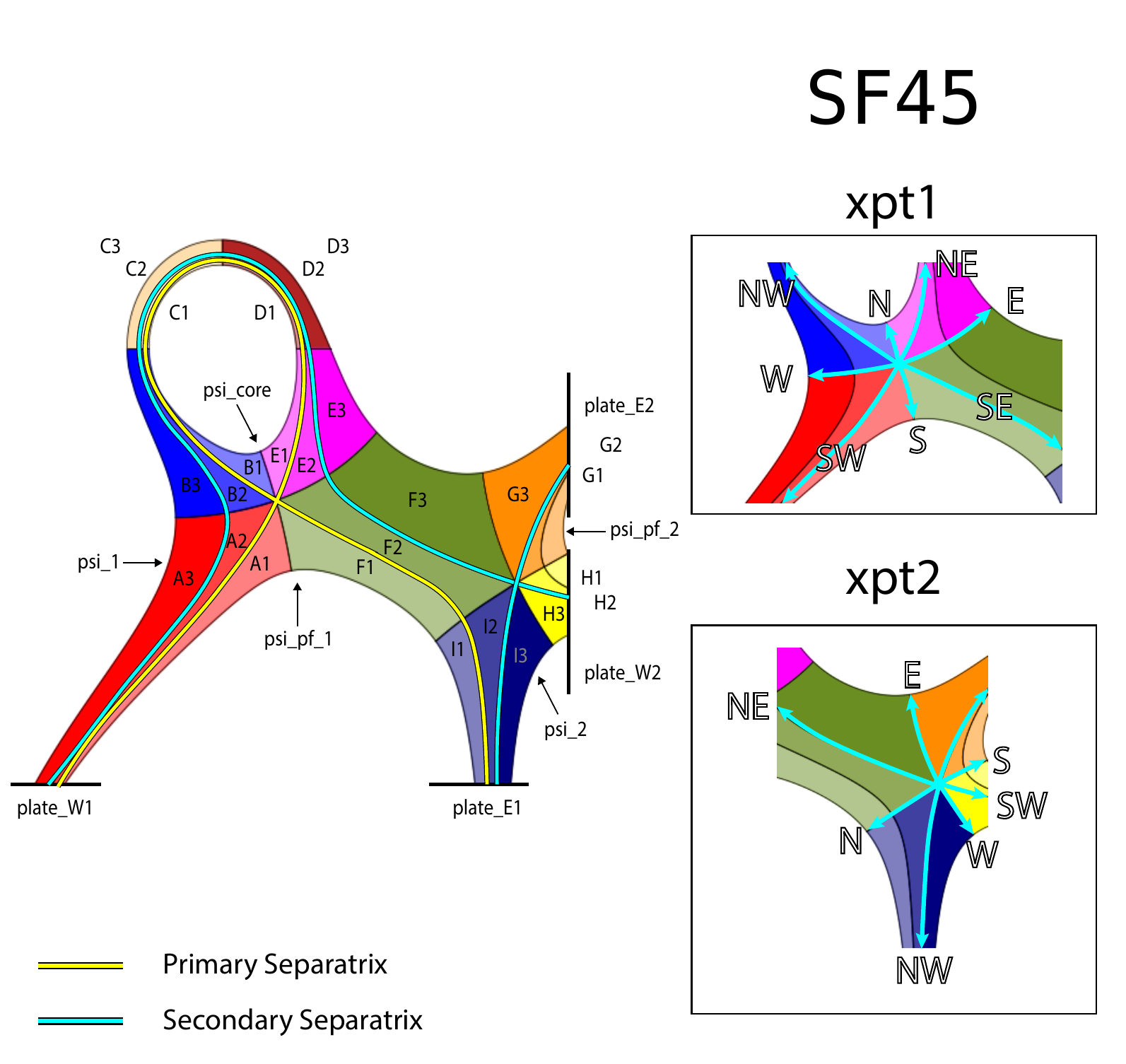}
        \caption{SF45 Patch-Map}
        \label{fig:sf45_patch_map}
\end{figure}
\begin{figure}[H]
    \centering
        \includegraphics[width=\textwidth]{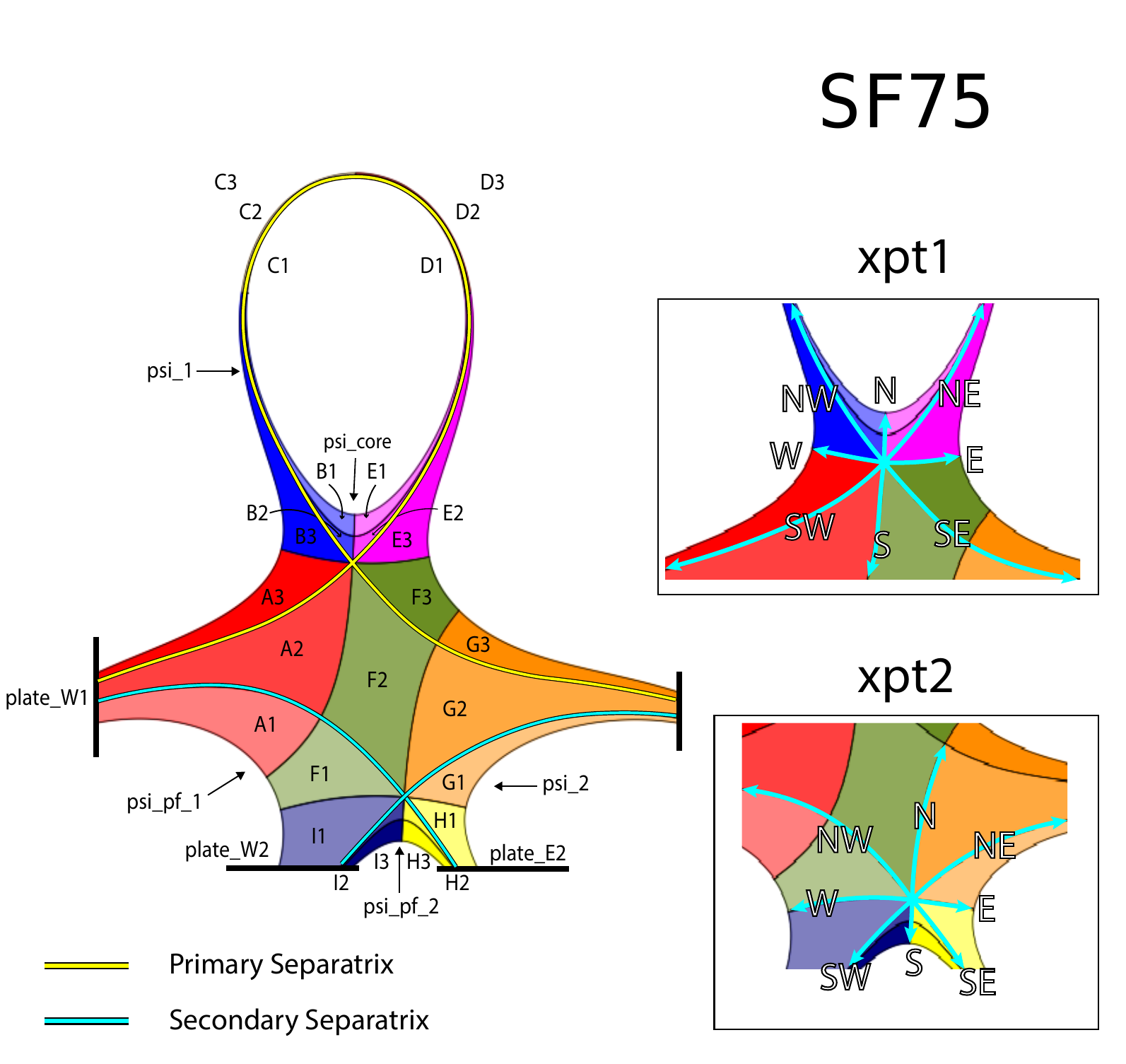}
        \caption{SF75 Patch-Map}
        \label{fig:sf75_patch_map}
\end{figure}
\begin{figure}[H]
    \centering
        \includegraphics[width=\textwidth]{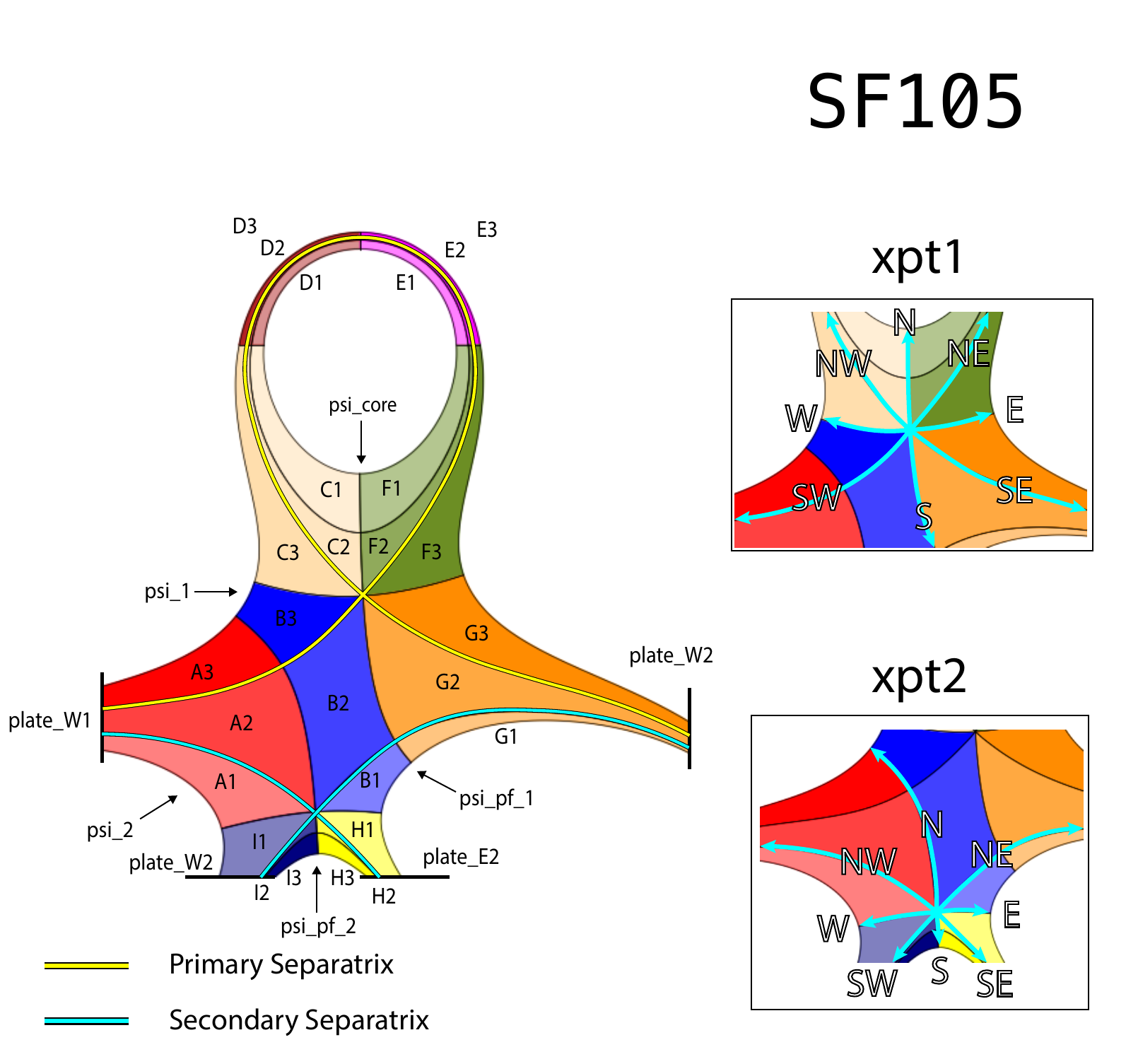}
        \caption{SF105 Patch-Map}
        \label{fig:sf105_patch_map}
\end{figure}
\begin{figure}[H]
    \centering
        \includegraphics[width=1.05\textwidth]{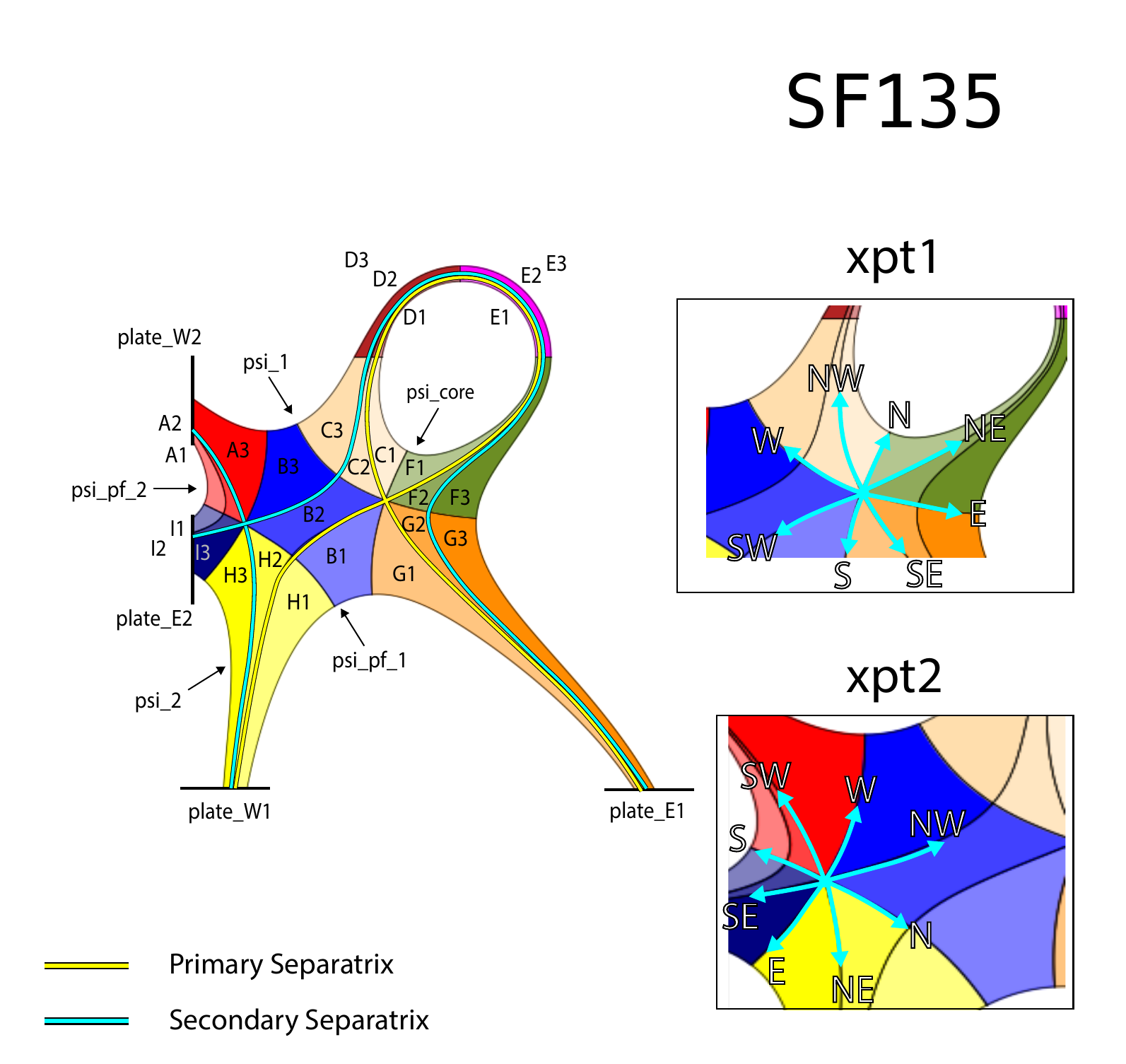}
        \caption{SF135 Patch-Map}
        \label{fig:sf135_patch_map}
\end{figure}
\begin{figure}[H]
    \centering
        \includegraphics[width=\textwidth]{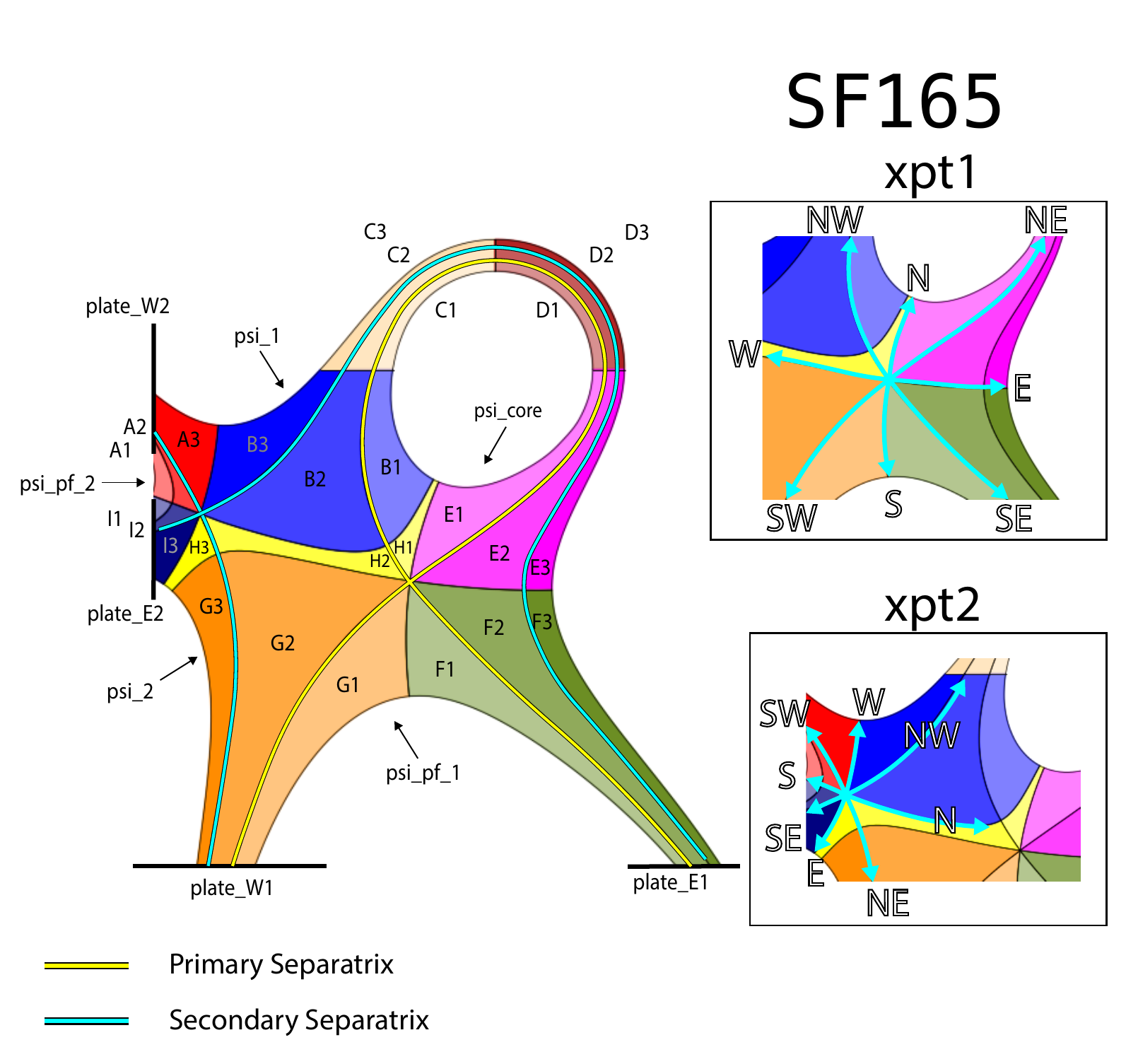}
        \caption{SF165 Patch-Map}
        \label{fig:sf165_patch_map}
\end{figure}

\begin{figure}[H]
    \centering
    \includegraphics[width=0.5\linewidth]{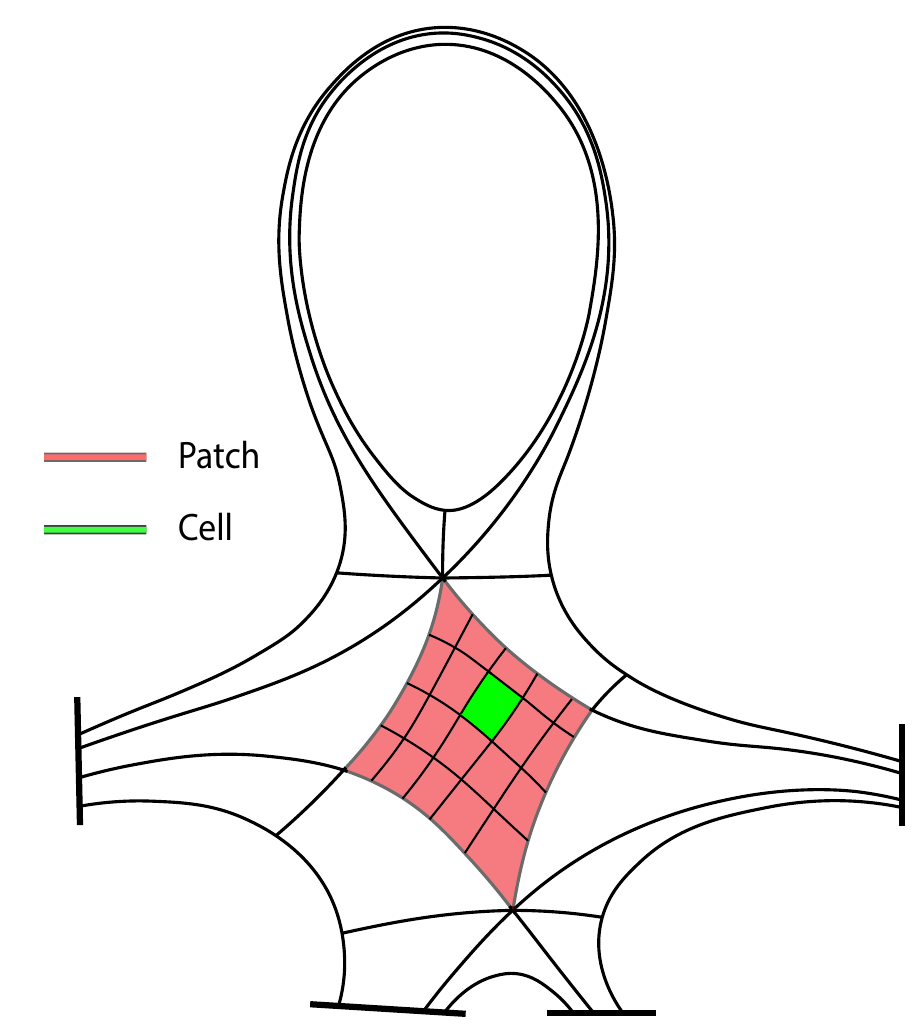}  
    \caption{INGRID Patch-Map; a subgrid is shown on one of the patches}
    \label{fig:patchmap_subgrid}
\end{figure}





\begin{figure}[H]
    \centering
    \includegraphics[width=\linewidth]{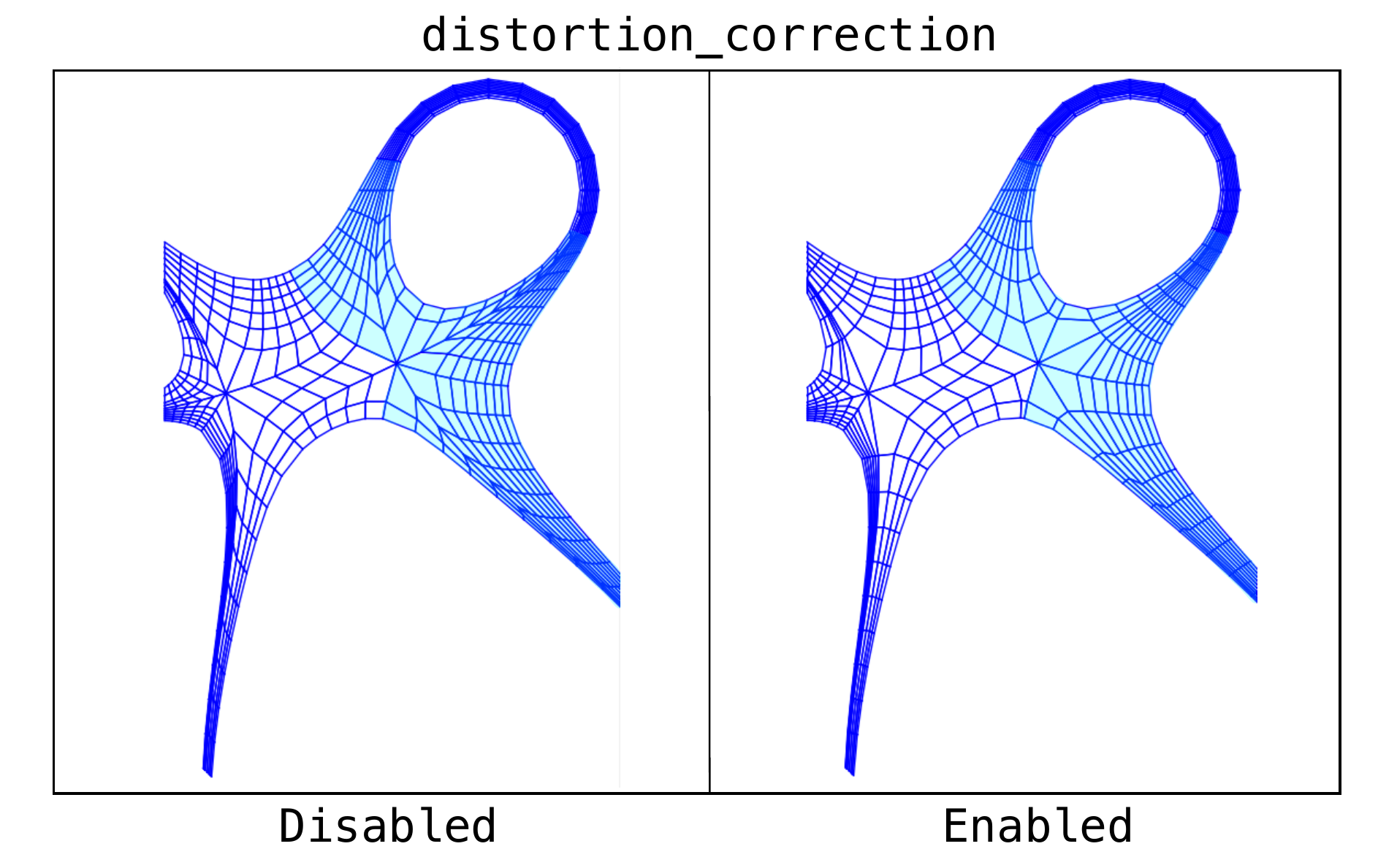}
    \caption{Comparison of SF135 grids generated with and without activation of the distortion\_correction tool. Highlighted regions illustrate regions of notably improved grid orthogonality.}
    \label{fig:distortion_correction}
\end{figure}

\begin{figure}[H]
    \centering
    \includegraphics[width=\linewidth]{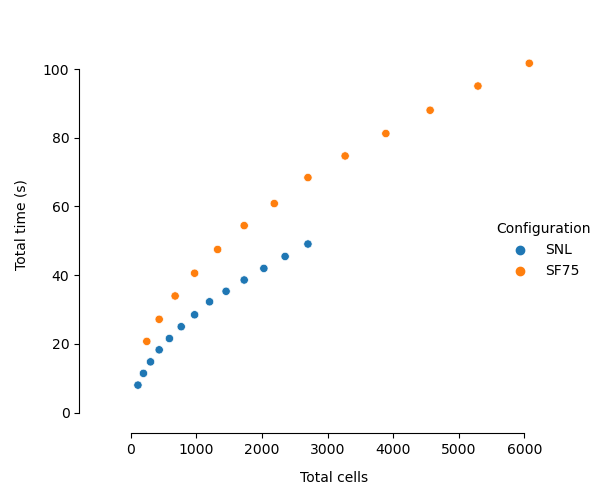}
    \caption{Scaling of grid generation follows a sublinear trend independent of configuration. Grids were generated with $n\times n$ many cells per Patch with $n = \{3, 4, 5, \dots, 15\}$. With $n \times n$ subgrid dimensions, SNL configurations contain $12n^2$ many cells, whereas SF75 configurations contain $27n^2$ many cells.}
    \label{fig:benchmark_grid_scaling}
\end{figure}

\begin{figure}[H]
    \centering
        \includegraphics[width=0.9\textwidth]{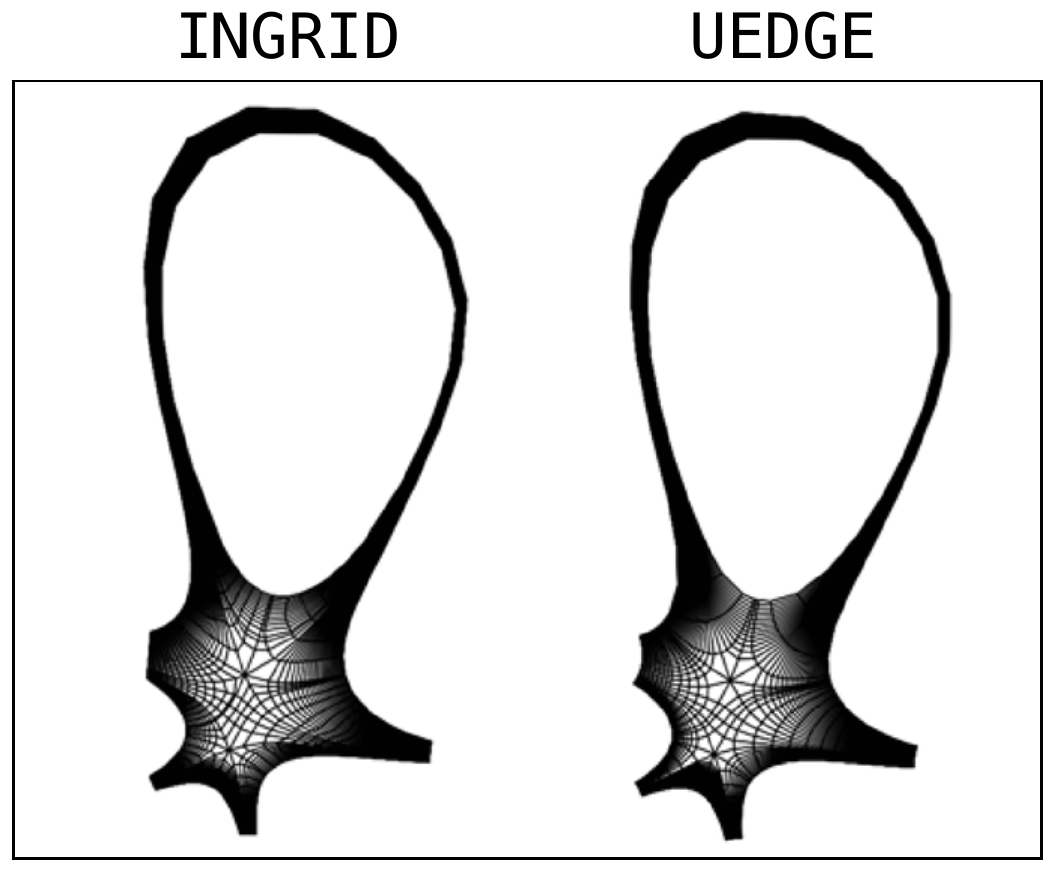}
        \caption{INGRID and UEDGE generated grids used for the benchmark calculations.}
        \label{fig:ingrid_grid}
\end{figure}

\begin{figure}[H]
    \centering
        \includegraphics[width=\textwidth]{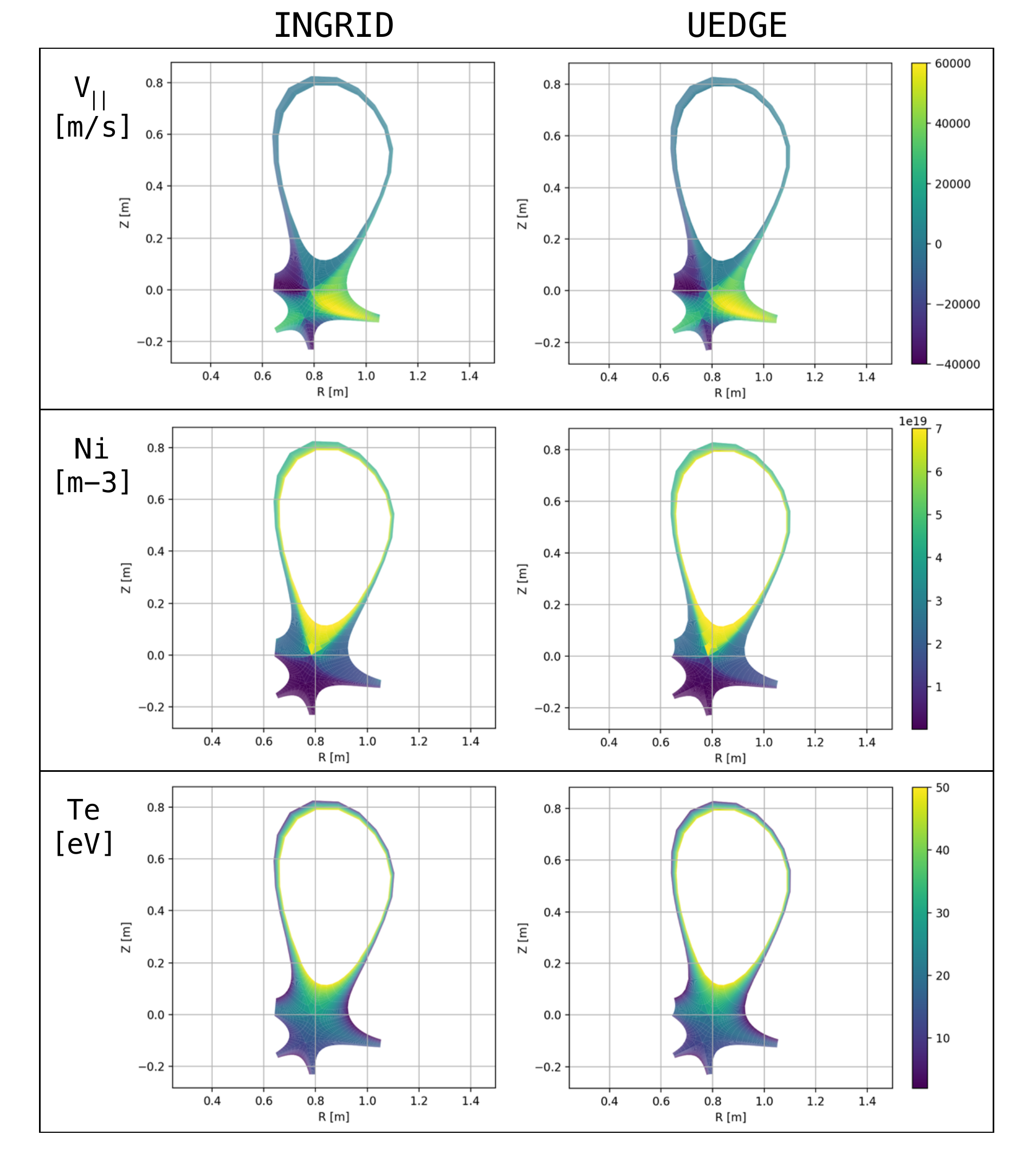}
        \caption{Results of the benchmark calculations run on INGRID and UEDGE generated grids.}
        \label{fig:benchmark_collection}
\end{figure}

\begin{figure}[H]
    \centering
        \includegraphics[width=0.9\textwidth]{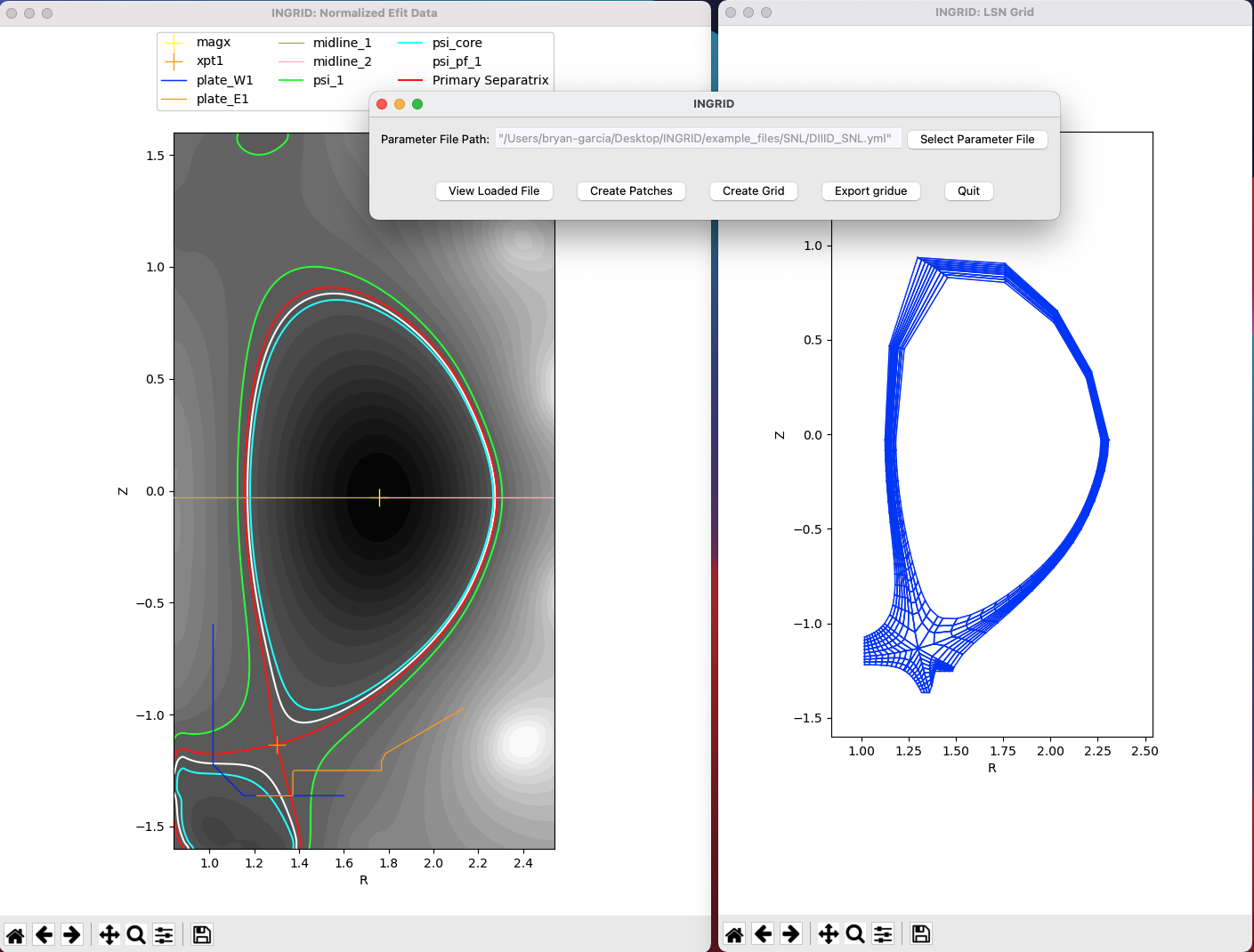}
        \caption{The INGRID GUI with loaded EFIT data and an INGRID generated grid plotted in separate windows. This interface allows users to load a parameter file, plot the contents, create a Patch map, create a grid, and export a gridue.}
        \label{fig:ingrid_gui}
\end{figure}

%% file: main.bbl
\providecommand{\noopsort}[1]{}\providecommand{\singleletter}[1]{#1}%
\begin{thebibliography}{26}%
\makeatletter
\providecommand \@ifxundefined [1]{%
 \@ifx{#1\undefined}
}%
\providecommand \@ifnum [1]{%
 \ifnum #1\expandafter \@firstoftwo
 \else \expandafter \@secondoftwo
 \fi
}%
\providecommand \@ifx [1]{%
 \ifx #1\expandafter \@firstoftwo
 \else \expandafter \@secondoftwo
 \fi
}%
\providecommand \natexlab [1]{#1}%
\providecommand \enquote  [1]{``#1''}%
\providecommand \bibnamefont  [1]{#1}%
\providecommand \bibfnamefont [1]{#1}%
\providecommand \citenamefont [1]{#1}%
\providecommand \href@noop [0]{\@secondoftwo}%
\providecommand \href [0]{\begingroup \@sanitize@url \@href}%
\providecommand \@href[1]{\@@startlink{#1}\@@href}%
\providecommand \@@href[1]{\endgroup#1\@@endlink}%
\providecommand \@sanitize@url [0]{\catcode `\\12\catcode `\$12\catcode
  `\&12\catcode `\#12\catcode `\^12\catcode `\_12\catcode `\%12\relax}%
\providecommand \@@startlink[1]{}%
\providecommand \@@endlink[0]{}%
\providecommand \url  [0]{\begingroup\@sanitize@url \@url }%
\providecommand \@url [1]{\endgroup\@href {#1}{\urlprefix }}%
\providecommand \urlprefix  [0]{URL }%
\providecommand \Eprint [0]{\href }%
\providecommand \doibase [0]{https://doi.org/}%
\providecommand \selectlanguage [0]{\@gobble}%
\providecommand \bibinfo  [0]{\@secondoftwo}%
\providecommand \bibfield  [0]{\@secondoftwo}%
\providecommand \translation [1]{[#1]}%
\providecommand \BibitemOpen [0]{}%
\providecommand \bibitemStop [0]{}%
\providecommand \bibitemNoStop [0]{.\EOS\space}%
\providecommand \EOS [0]{\spacefactor3000\relax}%
\providecommand \BibitemShut  [1]{\csname bibitem#1\endcsname}%
\let\auto@bib@innerbib\@empty
\bibitem [{\citenamefont {Kotschenreuther}(2007)}]{Kotschenreuther2007}%
  \BibitemOpen
  \bibfield  {author} {\bibinfo {author} {\bibfnamefont {M.}~\bibnamefont
  {Kotschenreuther}},\ }\href@noop {} {\bibfield  {journal} {\bibinfo
  {journal} {Physics of Plasmas}\ }\textbf {\bibinfo {volume} {14}},\ \bibinfo
  {pages} {072502} (\bibinfo {year} {2007})}\BibitemShut {NoStop}%
\bibitem [{\citenamefont {Ryutov}(2007)}]{Ryutov2007}%
  \BibitemOpen
  \bibfield  {author} {\bibinfo {author} {\bibfnamefont {D.~D.}\ \bibnamefont
  {Ryutov}},\ }\href@noop {} {\bibfield  {journal} {\bibinfo  {journal}
  {Physics of Plasmas}\ }\textbf {\bibinfo {volume} {14}},\ \bibinfo {pages}
  {064502} (\bibinfo {year} {2007})}\BibitemShut {NoStop}%
\bibitem [{\citenamefont {Ryutov}(2008)}]{Ryutov2008}%
  \BibitemOpen
  \bibfield  {author} {\bibinfo {author} {\bibfnamefont {D.~D.}\ \bibnamefont
  {Ryutov}},\ }\href@noop {} {\bibfield  {journal} {\bibinfo  {journal}
  {Physics of Plasmas}\ }\textbf {\bibinfo {volume} {15}},\ \bibinfo {pages}
  {092501} (\bibinfo {year} {2008})}\BibitemShut {NoStop}%
\bibitem [{\citenamefont {LaBombard}(2015)}]{LaBombard2015}%
  \BibitemOpen
  \bibfield  {author} {\bibinfo {author} {\bibfnamefont {B.}~\bibnamefont
  {LaBombard}},\ }\href@noop {} {\bibfield  {journal} {\bibinfo  {journal}
  {Nuclear Fusion}\ }\textbf {\bibinfo {volume} {55}},\ \bibinfo {pages}
  {053020} (\bibinfo {year} {2015})}\BibitemShut {NoStop}%
\bibitem [{\citenamefont {Rognlien}\ \emph {et~al.}(1999)\citenamefont
  {Rognlien}, \citenamefont {Ryutov}, \citenamefont {Mattor},\ and\
  \citenamefont {Porter}}]{Rognlien1999}%
  \BibitemOpen
  \bibfield  {author} {\bibinfo {author} {\bibfnamefont {T.~D.}\ \bibnamefont
  {Rognlien}}, \bibinfo {author} {\bibfnamefont {D.~D.}\ \bibnamefont
  {Ryutov}}, \bibinfo {author} {\bibfnamefont {N.}~\bibnamefont {Mattor}},\
  and\ \bibinfo {author} {\bibfnamefont {G.~D.}\ \bibnamefont {Porter}},\
  }\href@noop {} {\bibfield  {journal} {\bibinfo  {journal} {Physics of
  Plasmas}\ }\textbf {\bibinfo {volume} {6}},\ \bibinfo {pages} {1851}
  (\bibinfo {year} {1999})}\BibitemShut {NoStop}%
\bibitem [{\citenamefont {Wiesen}(2015)}]{Wiesen2015}%
  \BibitemOpen
  \bibfield  {author} {\bibinfo {author} {\bibfnamefont {S.}~\bibnamefont
  {Wiesen}},\ }\href@noop {} {\bibfield  {journal} {\bibinfo  {journal}
  {Journal of Nuclear Materials}\ }\textbf {\bibinfo {volume} {463}},\ \bibinfo
  {pages} {480} (\bibinfo {year} {2015})}\BibitemShut {NoStop}%
\bibitem [{\citenamefont {Simonini}(1994)}]{Simonini1994}%
  \BibitemOpen
  \bibfield  {author} {\bibinfo {author} {\bibfnamefont {R.}~\bibnamefont
  {Simonini}},\ }\href@noop {} {\bibfield  {journal} {\bibinfo  {journal}
  {Journal of Nuclear Materials}\ }\textbf {\bibinfo {volume} {34}},\ \bibinfo
  {pages} {368} (\bibinfo {year} {1994})}\BibitemShut {NoStop}%
\bibitem [{\citenamefont {Umansky}, \citenamefont {Day},\ and\ \citenamefont
  {Rognlien}(2005)}]{Umansky2005}%
  \BibitemOpen
  \bibfield  {author} {\bibinfo {author} {\bibfnamefont {M.~V.}\ \bibnamefont
  {Umansky}}, \bibinfo {author} {\bibfnamefont {M.~S.}\ \bibnamefont {Day}},\
  and\ \bibinfo {author} {\bibfnamefont {T.~D.}\ \bibnamefont {Rognlien}},\
  }\href@noop {} {\bibfield  {journal} {\bibinfo  {journal} {Numerical Heat
  Transfer, Part B}\ }\textbf {\bibinfo {volume} {47}},\ \bibinfo {pages} {533}
  (\bibinfo {year} {2005})}\BibitemShut {NoStop}%
\bibitem [{\citenamefont {Marchand}\ and\ \citenamefont
  {Dumberry}(1996)}]{Marchand1996}%
  \BibitemOpen
  \bibfield  {author} {\bibinfo {author} {\bibfnamefont {R.}~\bibnamefont
  {Marchand}}\ and\ \bibinfo {author} {\bibfnamefont {M.}~\bibnamefont
  {Dumberry}},\ }\bibfield  {title} {\enquote {\bibinfo {title} {{CARRE}: a
  quasi-orthogonal mesh generator for 2d edge plasma modelling},}\ }\href
  {https://doi.org/https://doi.org/10.1016/0010-4655(96)00052-5} {\bibfield
  {journal} {\bibinfo  {journal} {Computer Physics Communications}\ }\textbf
  {\bibinfo {volume} {96}},\ \bibinfo {pages} {232 -- 246} (\bibinfo {year}
  {1996})}\BibitemShut {NoStop}%
\bibitem [{\citenamefont {Taroni}(1992)}]{Taroni1992}%
  \BibitemOpen
  \bibfield  {author} {\bibinfo {author} {\bibfnamefont {A.}~\bibnamefont
  {Taroni}},\ }\bibfield  {title} {\enquote {\bibinfo {title} {The
  multi‐fluid codes {Edgeid} and {Edge2D}: Models and results},}\ }\href@noop
  {} {\bibfield  {journal} {\bibinfo  {journal} {Contribution to Plasma
  Physics}\ }\textbf {\bibinfo {volume} {34}},\ \bibinfo {pages} {448}
  (\bibinfo {year} {1992})}\BibitemShut {NoStop}%
\bibitem [{\citenamefont {Izacard}\ and\ \citenamefont
  {Umansky}(2017)}]{Izacard2017}%
  \BibitemOpen
  \bibfield  {author} {\bibinfo {author} {\bibfnamefont {O.}~\bibnamefont
  {Izacard}}\ and\ \bibinfo {author} {\bibfnamefont {M.}~\bibnamefont
  {Umansky}},\ }\href@noop {} {\enquote {\bibinfo {title} {Gingred, a general
  grid generator for 2d edge plasma modeling},}\ } (\bibinfo {year} {2017}),\
  \Eprint {https://arxiv.org/abs/1705.05717} {arXiv:1705.05717
  [physics.plasm-ph]} \BibitemShut {NoStop}%
\bibitem [{\citenamefont {Freidberg}(2014)}]{Freidberg2014}%
  \BibitemOpen
  \bibfield  {author} {\bibinfo {author} {\bibfnamefont {J.~P.}\ \bibnamefont
  {Freidberg}},\ }\href@noop {} {\emph {\bibinfo {title} {Ideal MHD}}}\
  (\bibinfo  {publisher} {Cambridge University Press},\ \bibinfo {address} {New
  York},\ \bibinfo {year} {2014})\BibitemShut {NoStop}%
\bibitem [{\citenamefont {Ryutov}, \citenamefont {Makowski},\ and\
  \citenamefont {Umansky}(2010)}]{Ryutov2010}%
  \BibitemOpen
  \bibfield  {author} {\bibinfo {author} {\bibfnamefont {D.~D.}\ \bibnamefont
  {Ryutov}}, \bibinfo {author} {\bibfnamefont {M.~A.}\ \bibnamefont
  {Makowski}},\ and\ \bibinfo {author} {\bibfnamefont {M.~V.}\ \bibnamefont
  {Umansky}},\ }\bibfield  {title} {\enquote {\bibinfo {title} {Local
  properties of the magnetic field in a snowflake divertor},}\ }\href
  {https://doi.org/10.1088/0741-3335/52/10/105001} {\bibfield  {journal}
  {\bibinfo  {journal} {Plasma Physics and Controlled Fusion}\ }\textbf
  {\bibinfo {volume} {52}},\ \bibinfo {pages} {105001} (\bibinfo {year}
  {2010})}\BibitemShut {NoStop}%
\bibitem [{\citenamefont {Lao}\ \emph {et~al.}(1985)\citenamefont {Lao},
  \citenamefont {John}, \citenamefont {Stambaugh}, \citenamefont {Kellman},\
  and\ \citenamefont {Pfeiffer}}]{Lao_1985}%
  \BibitemOpen
  \bibfield  {author} {\bibinfo {author} {\bibfnamefont {L.}~\bibnamefont
  {Lao}}, \bibinfo {author} {\bibfnamefont {H.~S.}\ \bibnamefont {John}},
  \bibinfo {author} {\bibfnamefont {R.}~\bibnamefont {Stambaugh}}, \bibinfo
  {author} {\bibfnamefont {A.}~\bibnamefont {Kellman}},\ and\ \bibinfo {author}
  {\bibfnamefont {W.}~\bibnamefont {Pfeiffer}},\ }\bibfield  {title} {\enquote
  {\bibinfo {title} {Reconstruction of current profile parameters and plasma
  shapes in tokamaks},}\ }\href {https://doi.org/10.1088/0029-5515/25/11/007}
  {\bibfield  {journal} {\bibinfo  {journal} {Nuclear Fusion}\ }\textbf
  {\bibinfo {volume} {25}},\ \bibinfo {pages} {1611--1622} (\bibinfo {year}
  {1985})}\BibitemShut {NoStop}%
\bibitem [{\citenamefont {Press}\ \emph {et~al.}(1992)\citenamefont {Press},
  \citenamefont {Teukolsky}, \citenamefont {Vetterling},\ and\ \citenamefont
  {Flannery}}]{Press_1992}%
  \BibitemOpen
  \bibfield  {author} {\bibinfo {author} {\bibfnamefont {W.~H.}\ \bibnamefont
  {Press}}, \bibinfo {author} {\bibfnamefont {S.~A.}\ \bibnamefont
  {Teukolsky}}, \bibinfo {author} {\bibfnamefont {W.~T.}\ \bibnamefont
  {Vetterling}},\ and\ \bibinfo {author} {\bibfnamefont {B.~P.}\ \bibnamefont
  {Flannery}},\ }\href@noop {} {\emph {\bibinfo {title} {Numerical Recipes in
  C}}},\ \bibinfo {edition} {2nd}\ ed.\ (\bibinfo  {publisher} {Cambridge
  University Press},\ \bibinfo {address} {Cambridge, USA},\ \bibinfo {year}
  {1992})\BibitemShut {NoStop}%
\bibitem [{\citenamefont {Virtanen}\ \emph {et~al.}(2019)\citenamefont
  {Virtanen}, \citenamefont {Gommers}, \citenamefont {Oliphant}, \citenamefont
  {Haberland}, \citenamefont {Reddy}, \citenamefont {Cournapeau}, \citenamefont
  {Burovski}, \citenamefont {Peterson}, \citenamefont {Weckesser},
  \citenamefont {Bright}, \citenamefont {van~der Walt}, \citenamefont {Brett},
  \citenamefont {Wilson}, \citenamefont {Millman}, \citenamefont {Mayorov},
  \citenamefont {Nelson}, \citenamefont {Jones}, \citenamefont {Kern},
  \citenamefont {Larson}, \citenamefont {Carey}, \citenamefont {İlhan Polat},
  \citenamefont {Feng}, \citenamefont {Moore}, \citenamefont {VanderPlas},
  \citenamefont {Laxalde}, \citenamefont {Perktold}, \citenamefont {Cimrman},
  \citenamefont {Henriksen}, \citenamefont {Quintero}, \citenamefont {Harris},
  \citenamefont {Archibald}, \citenamefont {Ribeiro}, \citenamefont
  {Pedregosa}, \citenamefont {van Mulbregt},\ and\ \citenamefont
  {Contributors}}]{virtanen2019scipy}%
  \BibitemOpen
  \bibfield  {author} {\bibinfo {author} {\bibfnamefont {P.}~\bibnamefont
  {Virtanen}}, \bibinfo {author} {\bibfnamefont {R.}~\bibnamefont {Gommers}},
  \bibinfo {author} {\bibfnamefont {T.~E.}\ \bibnamefont {Oliphant}}, \bibinfo
  {author} {\bibfnamefont {M.}~\bibnamefont {Haberland}}, \bibinfo {author}
  {\bibfnamefont {T.}~\bibnamefont {Reddy}}, \bibinfo {author} {\bibfnamefont
  {D.}~\bibnamefont {Cournapeau}}, \bibinfo {author} {\bibfnamefont
  {E.}~\bibnamefont {Burovski}}, \bibinfo {author} {\bibfnamefont
  {P.}~\bibnamefont {Peterson}}, \bibinfo {author} {\bibfnamefont
  {W.}~\bibnamefont {Weckesser}}, \bibinfo {author} {\bibfnamefont
  {J.}~\bibnamefont {Bright}}, \bibinfo {author} {\bibfnamefont {S.~J.}\
  \bibnamefont {van~der Walt}}, \bibinfo {author} {\bibfnamefont
  {M.}~\bibnamefont {Brett}}, \bibinfo {author} {\bibfnamefont
  {J.}~\bibnamefont {Wilson}}, \bibinfo {author} {\bibfnamefont {K.~J.}\
  \bibnamefont {Millman}}, \bibinfo {author} {\bibfnamefont {N.}~\bibnamefont
  {Mayorov}}, \bibinfo {author} {\bibfnamefont {A.~R.~J.}\ \bibnamefont
  {Nelson}}, \bibinfo {author} {\bibfnamefont {E.}~\bibnamefont {Jones}},
  \bibinfo {author} {\bibfnamefont {R.}~\bibnamefont {Kern}}, \bibinfo {author}
  {\bibfnamefont {E.}~\bibnamefont {Larson}}, \bibinfo {author} {\bibfnamefont
  {C.}~\bibnamefont {Carey}}, \bibinfo {author} {\bibnamefont {İlhan Polat}},
  \bibinfo {author} {\bibfnamefont {Y.}~\bibnamefont {Feng}}, \bibinfo {author}
  {\bibfnamefont {E.~W.}\ \bibnamefont {Moore}}, \bibinfo {author}
  {\bibfnamefont {J.}~\bibnamefont {VanderPlas}}, \bibinfo {author}
  {\bibfnamefont {D.}~\bibnamefont {Laxalde}}, \bibinfo {author} {\bibfnamefont
  {J.}~\bibnamefont {Perktold}}, \bibinfo {author} {\bibfnamefont
  {R.}~\bibnamefont {Cimrman}}, \bibinfo {author} {\bibfnamefont
  {I.}~\bibnamefont {Henriksen}}, \bibinfo {author} {\bibfnamefont {E.~A.}\
  \bibnamefont {Quintero}}, \bibinfo {author} {\bibfnamefont {C.~R.}\
  \bibnamefont {Harris}}, \bibinfo {author} {\bibfnamefont {A.~M.}\
  \bibnamefont {Archibald}}, \bibinfo {author} {\bibfnamefont {A.~H.}\
  \bibnamefont {Ribeiro}}, \bibinfo {author} {\bibfnamefont {F.}~\bibnamefont
  {Pedregosa}}, \bibinfo {author} {\bibfnamefont {P.}~\bibnamefont {van
  Mulbregt}},\ and\ \bibinfo {author} {\bibfnamefont {S.~.~.}\ \bibnamefont
  {Contributors}},\ }\href@noop {} {\enquote {\bibinfo {title} {Scipy
  1.0--fundamental algorithms for scientific computing in python},}\ }
  (\bibinfo {year} {2019}),\ \Eprint {https://arxiv.org/abs/1907.10121}
  {arXiv:1907.10121 [cs.MS]} \BibitemShut {NoStop}%
\bibitem [{\citenamefont {Enthought}()}]{scipy.interpolate}%
  \BibitemOpen
  \bibfield  {author} {\bibinfo {author} {\bibfnamefont {I.}~\bibnamefont
  {Enthought}},\ }\href@noop {} {\enquote {\bibinfo {title} {{SciPy
  Interpolate} {GitHub repository}},}\ }\bibinfo {howpublished}
  {\url{https://github.com/scipy/scipy/tree/master/scipy/interpolate}}\BibitemShut
  {NoStop}%
\bibitem [{sci()}]{scipy.optimize.root_docs}%
  \BibitemOpen
  \href@noop {} {\enquote {\bibinfo {title} {{scipy.optimize.root}
  documentation},}\ }\bibinfo {howpublished}
  {\url{https://docs.scipy.org/doc/scipy/reference/generated/scipy.optimize.root.html}}\BibitemShut
  {NoStop}%
\bibitem [{\citenamefont {Hindmarsh}(1992)}]{osti_145724}%
  \BibitemOpen
  \bibfield  {author} {\bibinfo {author} {\bibfnamefont {A.}~\bibnamefont
  {Hindmarsh}},\ }\bibfield  {title} {\enquote {\bibinfo {title} {Odepack. a
  collection of ode system solvers},}\ }\href
  {https://www.osti.gov/biblio/145724} {\  (\bibinfo {year}
  {1992})}\BibitemShut {NoStop}%
\bibitem [{\citenamefont {Meneghini}\ \emph {et~al.}(2015)\citenamefont
  {Meneghini}, \citenamefont {Smith}, \citenamefont {Lao}, \citenamefont
  {Izacard}, \citenamefont {Ren}, \citenamefont {Park}, \citenamefont {Candy},
  \citenamefont {Wang}, \citenamefont {Luna}, \citenamefont {Izzo},
  \citenamefont {Grierson}, \citenamefont {Snyder}, \citenamefont {Holland},
  \citenamefont {Penna}, \citenamefont {Lu}, \citenamefont {Raum},
  \citenamefont {McCubbin}, \citenamefont {Orlov}, \citenamefont {Belli},
  \citenamefont {Ferraro}, \citenamefont {Prater}, \citenamefont {Osborne},
  \citenamefont {Turnbull},\ and\ \citenamefont {Staebler}}]{Meneghini_2015}%
  \BibitemOpen
  \bibfield  {author} {\bibinfo {author} {\bibfnamefont {O.}~\bibnamefont
  {Meneghini}}, \bibinfo {author} {\bibfnamefont {S.}~\bibnamefont {Smith}},
  \bibinfo {author} {\bibfnamefont {L.}~\bibnamefont {Lao}}, \bibinfo {author}
  {\bibfnamefont {O.}~\bibnamefont {Izacard}}, \bibinfo {author} {\bibfnamefont
  {Q.}~\bibnamefont {Ren}}, \bibinfo {author} {\bibfnamefont {J.}~\bibnamefont
  {Park}}, \bibinfo {author} {\bibfnamefont {J.}~\bibnamefont {Candy}},
  \bibinfo {author} {\bibfnamefont {Z.}~\bibnamefont {Wang}}, \bibinfo {author}
  {\bibfnamefont {C.}~\bibnamefont {Luna}}, \bibinfo {author} {\bibfnamefont
  {V.}~\bibnamefont {Izzo}}, \bibinfo {author} {\bibfnamefont {B.}~\bibnamefont
  {Grierson}}, \bibinfo {author} {\bibfnamefont {P.}~\bibnamefont {Snyder}},
  \bibinfo {author} {\bibfnamefont {C.}~\bibnamefont {Holland}}, \bibinfo
  {author} {\bibfnamefont {J.}~\bibnamefont {Penna}}, \bibinfo {author}
  {\bibfnamefont {G.}~\bibnamefont {Lu}}, \bibinfo {author} {\bibfnamefont
  {P.}~\bibnamefont {Raum}}, \bibinfo {author} {\bibfnamefont {A.}~\bibnamefont
  {McCubbin}}, \bibinfo {author} {\bibfnamefont {D.}~\bibnamefont {Orlov}},
  \bibinfo {author} {\bibfnamefont {E.}~\bibnamefont {Belli}}, \bibinfo
  {author} {\bibfnamefont {N.}~\bibnamefont {Ferraro}}, \bibinfo {author}
  {\bibfnamefont {R.}~\bibnamefont {Prater}}, \bibinfo {author} {\bibfnamefont
  {T.}~\bibnamefont {Osborne}}, \bibinfo {author} {\bibfnamefont
  {A.}~\bibnamefont {Turnbull}},\ and\ \bibinfo {author} {\bibfnamefont
  {G.}~\bibnamefont {Staebler}},\ }\bibfield  {title} {\enquote {\bibinfo
  {title} {Integrated modeling applications for tokamak experiments with
  {OMFIT}},}\ }\href {https://doi.org/10.1088/0029-5515/55/8/083008} {\bibfield
   {journal} {\bibinfo  {journal} {Nuclear Fusion}\ }\textbf {\bibinfo {volume}
  {55}},\ \bibinfo {pages} {083008} (\bibinfo {year} {2015})}\BibitemShut
  {NoStop}%
\bibitem [{\citenamefont {Meneghini}\ and\ \citenamefont
  {Lao}(2013)}]{Orso_MENEGHINI2013}%
  \BibitemOpen
  \bibfield  {author} {\bibinfo {author} {\bibfnamefont {O.}~\bibnamefont
  {Meneghini}}\ and\ \bibinfo {author} {\bibfnamefont {L.}~\bibnamefont
  {Lao}},\ }\bibfield  {title} {\enquote {\bibinfo {title} {Integrated modeling
  of tokamak experiments with omfit},}\ }\href
  {https://doi.org/10.1585/pfr.8.2403009} {\bibfield  {journal} {\bibinfo
  {journal} {Plasma and Fusion Research}\ }\textbf {\bibinfo {volume} {8}},\
  \bibinfo {pages} {2403009--2403009} (\bibinfo {year} {2013})}\BibitemShut
  {NoStop}%
\bibitem [{PyU()}]{PyUEDGE}%
  \BibitemOpen
  \href@noop {} {\enquote {\bibinfo {title} {{UEDGE} code {GitHub}
  repository},}\ }\bibinfo {howpublished} {https://github.com/LLNL/UEDGE},\
  \bibinfo {note} {accessed: 2021-01-18}\BibitemShut {NoStop}%
\bibitem [{\citenamefont {Simonov}(2006)}]{PyYAML}%
  \BibitemOpen
  \bibfield  {author} {\bibinfo {author} {\bibfnamefont {K.}~\bibnamefont
  {Simonov}},\ }\href@noop {} {\enquote {\bibinfo {title} {Pyyaml homepage},}\
  }\bibinfo {howpublished} {https://pyyaml.org/wiki/PyYAML} (\bibinfo {year}
  {2006})\BibitemShut {NoStop}%
\bibitem [{\citenamefont {Rensink}\ and\ \citenamefont
  {Rognlien}(2017)}]{Rensink2017}%
  \BibitemOpen
  \bibfield  {author} {\bibinfo {author} {\bibfnamefont {M.~E.}\ \bibnamefont
  {Rensink}}\ and\ \bibinfo {author} {\bibfnamefont {T.~D.}\ \bibnamefont
  {Rognlien}},\ }\href@noop {} {\enquote {\bibinfo {title} {Mapping of
  orthogonal 2d flux coordinates for two nearby magnetic x-points to logically
  rectangular domains},}\ }\bibinfo {type} {Tech. Rep.}\ \bibinfo {number}
  {LLNL-TR-731515}\ (\bibinfo  {institution} {Lawrence Livermore National
  Laboratory},\ \bibinfo {address} {Livermore, California},\ \bibinfo {year}
  {2017})\BibitemShut {NoStop}%
\bibitem [{\citenamefont {Garcia}\ \emph {et~al.}(2019)\citenamefont {Garcia},
  \citenamefont {Watkins}, \citenamefont {Guterl},\ and\ \citenamefont
  {Umansky}}]{ingrid_github}%
  \BibitemOpen
  \bibfield  {author} {\bibinfo {author} {\bibfnamefont {B.~M.}\ \bibnamefont
  {Garcia}}, \bibinfo {author} {\bibfnamefont {J.}~\bibnamefont {Watkins}},
  \bibinfo {author} {\bibfnamefont {J.}~\bibnamefont {Guterl}},\ and\ \bibinfo
  {author} {\bibfnamefont {M.~V.}\ \bibnamefont {Umansky}},\ }\href@noop {}
  {\enquote {\bibinfo {title} {{INGRID} code {GitHub repository}},}\ }\bibinfo
  {howpublished} {\url{https://github.com/LLNL/INGRID}} (\bibinfo {year}
  {2019})\BibitemShut {NoStop}%
\bibitem [{\citenamefont {{B. M. Garcia}}\ \emph {et~al.}(2020)\citenamefont
  {{B. M. Garcia}}, \citenamefont {{J. Watkins}}, \citenamefont {{J. Guterl}},\
  and\ \citenamefont {{M. V. Umansky}}}]{ingrid_readthedocs}%
  \BibitemOpen
  \bibfield  {author} {\bibinfo {author} {\bibnamefont {{B. M. Garcia}}},
  \bibinfo {author} {\bibnamefont {{J. Watkins}}}, \bibinfo {author}
  {\bibnamefont {{J. Guterl}}},\ and\ \bibinfo {author} {\bibnamefont {{M. V.
  Umansky}}},\ }\href@noop {} {\enquote {\bibinfo {title} {{INGRID Read the
  Docs}},}\ }\bibinfo {howpublished}
  {\url{https://ingrid.readthedocs.io/en/latest/}} (\bibinfo {year}
  {2020})\BibitemShut {NoStop}%
\end{thebibliography}%
